\documentclass[aps,twocolumn,prc,superscriptaddress,noshowpacs,nofootinbib,noshowkeys,floatfix]{revtex4}

\usepackage{epsfig}
\usepackage[colorlinks=true,linktocpage=true,linkcolor=blue,citecolor=blue]{hyperref}
\usepackage[usenames,dvipsnames]{color}
\usepackage{amsmath, amssymb}
\usepackage{multirow}
\usepackage{longtable}
\usepackage{color}
\usepackage[normalem]{ulem}  
\usepackage{float}

\renewcommand\sout{\bgroup \color{blue} \ULdepth=-.5ex \ULset}

\newcommand{\I}{{\rm I}}
\newcommand{\T}{{\rm T}}

\begin{document}
\title{Flow fluctuations and kinetic freeze-out of identified hadrons at energies available at the CERN Super Proton Synchrotron}
\author{Sudhir Pandurang Rode}
\affiliation{Veksler and Baldin Laboratory of High Energy Physics, Joint Institute for Nuclear Research, Dubna, 141980, Moscow region, Russian Federation}

\author{Partha Pratim Bhaduri}
\affiliation{Variable Energy Cyclotron Centre, HBNI, 1/AF Bidhan Nagar, Kolkata 700 064, India}

\author{Amaresh Jaiswal}
\affiliation{School of Physical Sciences, National Institute of Science Education and Research, An OCC of Homi Bhabha National Institute, Jatni-752050, Odisha, India}

\date{\today}

\begin{abstract}

We investigate the effect of flow fluctuations, incorporated in non boost-invariant blast-wave model, on kinetic freeze-out parameters of identified hadrons in low energy relativistic heavy-ion collisions. For the purpose of this study, we use the transverse momentum spectra of the identified hadrons produced in central Pb--Pb collisions, at SPS energies ranging from $\rm E_{Lab}=20A-158A $ GeV, and analyze them within a modified non boost-invariant blast wave model. We perform simultaneous fits of the transverse momentum spectra for light hadrons ($\pi^{-}$, $K^{\pm}$, $p$) and heavy strange hadrons ($\Lambda$, $\bar{\Lambda}$, $\phi$, $\Xi^{\pm}$, $\Omega^{\pm}$) separately. We also fit the transverse momentum spectra of charmonia ($J/\Psi$, $\Psi'$) at $\rm E_{Lab}=158A $ GeV. Our findings suggest that the inclusion of flow fluctuations enhances kinetic freeze-out temperature in case of light and heavy strange hadrons and reduces the corresponding transverse flow velocities. Moreover, we find that the kinetic freeze-out parameters of the charmonia at $\rm E_{Lab}=158A $ GeV are least affected by inclusion of flow fluctuations. Based on this, we make predictions which can provide further insights on the role of flow fluctuations in relativistic heavy-ion collisions.

\end{abstract}

\maketitle


\section{Introduction}
\label{}

Collisions of relativistically accelerated heavy-ions in the laboratory allow production and study of hot and dense Quantum Chromo-Dynamics (QCD) matter~\cite{Florkowski:2014yza, UWHeinz, BraunMunzinger:2008tz}. Tuning of collision energy can enable the possibility of creating nuclear matter at various temperatures and baryon densities which can probe a large part of QCD phase diagram. Relativistic Heavy Ion Collider (RHIC)~\cite{rhic1,rhic2} and Large Hadron Collider (LHC)~\cite{lhc1,lhc2,lhc3} accelerate nuclei with ultra-relativistic speeds which creates medium having thermodynamic conditions of high temperatures and negligible baryon chemical potentials. Lattice QCD (lQCD) simulations~\cite{Fodor:2009ax, Bazavov:2009zn, Borsanyi:2010cj, Bazavov:2014pvz, Borsanyi:2013bia} are well suited for the study of such medium.

Nuclear matter corresponding to the region of moderate temperature and finite net baryon densities in QCD phase diagram is created by lowering the beam energies. The application of lQCD to study such matter is limited. However, in recent times, the interest in studying nuclear collisions at these energies has been rejuvenated and many ongoing as well as upcoming accelerator facilities at RHIC~\cite{Bzdak:2019pkr}, Super Proton Synchrotron (SPS)~\cite{Lewicki:2020mqr,Agnello:2018evr}, Facility for Anti-proton Ion Research (FAIR~\cite{CBM-physics,Senger:2019ccc} and Nuclotron-based Ion Collider fAcility (NICA)~\cite{nica}, have performed and planned various experimental programs. This includes the beam energy scan (BES) and STAR Fxt (fixed target) program of RHIC, NA61 and NA60+ experiments at SPS, Compressed Baryonic Matter (CBM) experiment at FAIR, and Baryonic Matter at the Nuclotron (BM@N) and Multi-Purpose Detector (MPD) experiment at NICA. The systematic interpretation of the available data from earlier fixed-target mode experiments at AGS and SPS facilities in these beam energy ranges can allow an appropriate utilization of the upcoming facilities. Out of several challenges, estimation of freeze-out conditions of the fireball at various beam energies has been one of the compelling topics in heavy-ion collisions.

The particle chemistry of the fireball stabilizes during the chemical freeze-out as the inelastic scatterings stop, whereas, during kinetic freeze-out, the momentum distributions of the hadrons are frozen. The quark flavour dependent multiple chemical freeze-out scenario where strange hadrons fix their composition earlier than light hadrons, was predicted by the authors of Ref.~\cite{Chatterjee:2013yga}. Similar observations were found for mass dependent kinetic freeze-out of the measured hadrons in the fixed target energy domain~\cite{Rode:2020vhu}. In general, hydro-inspired blast-wave model can be used to describe kinetic freeze-out conditions~\cite{Florkowski:2004tn}. The particle spectra from hydrodynamics was described by assuming the emission from cylindrically symmetric and boost-invariant fireball~\cite{Schnedermann}. Over the years there have been several modifications to the original formulation of the blast wave model. Recently, the formulation of non-boost-invariant blast-wave model~\cite{Dobler:1999ju} was employed at AGS and SPS energies in our previous works to describe the transverse and longitudinal spectra of identified hadrons~\cite{Rode:2018hlj,Rode:2020vhu}.

The main assumptions in the formulation of the blast wave model are the following: the freeze-out isotherm is described at a constant proper time ($\tau =$ const) and the transverse rapidity profile at the isotherm has a linear form. Other assumptions are neglecting the presence of flow fluctuations, on-mass shell distributions functions, a homogeneous number density and absence of resonance feed-down. The assumption of absence of resonance feed-down has been taken into consideration by us for pions in our previous work~\cite{Rode:2018hlj}. In the present article, we have accounted for flow fluctuations following the Ref.~\cite{Akkelin:2009nx} which was applied to the boost-invariant blast-wave model. Due to the finite size of the systems generated in heavy-ion collisions, significant fluctuations are expected during the initial stages of the collisions. This applies even to collisions with fixed impact parameters. The initial conditions of hydrodynamical calculations are sensitive to these fluctuations whose impact can survive till the freeze-out. Therefore it is crucial to consider these fluctuations in the blast wave model. Further details regarding this implementation will be presented in the following section.

In this article, we have modified the non-boost-invariant blast-wave model following the idea from Ref.~\cite{Akkelin:2009nx} and employed this modified non boost invariant blast-wave model to study the effect of flow fluctuations on the kinetic freeze-out conditions of identified hadrons in central Pb--Pb collisions at SPS energies. In Ref.~\cite{Akkelin:2009nx}, the authors have considered two different formulations, namely, Flat or Uniform and Gaussian distribution of hydrodynamical velocities for implementing the flow fluctuations (more details in sec. II). To accomplish our goal, we examine the $p_{\rm T}$-spectra of identified particles within beam energy range $\rm E_{Lab}=20A-158 $A GeV. The identified particles are catogorized according to their mass as, light hadrons ($\pi^{-}$, $K^{\pm}$, $p$) and heavy strange hadrons ($\Lambda$, $\bar{\Lambda}$, $\phi$, $\Xi^{\pm}$, $\Omega^{\pm}$) as well as charmonia ($J/\psi$ and $\psi^{'}$ only at $\rm E_{Lab}=158 $A GeV). The rapidity spectra are not analyzed in this article since it is expected to be insensitive to the changes in the transverse flow profile\footnotemark \footnotetext[1]{We have explicitly verified that the rapidity distributions are insensitive to the incorporation of fluctuations in the transverse flow profile.}. Our findings in this article predicts higher kinetic freeze-out temperature and lower in transverse flow velocity using both uniform as well as Gaussian formulations compared to no fluctuations scenario for both light as well as heavy strange hadrons across all analyzed beam energies. Interestingly, the kinetic freeze-out temperature and transverse flow velocity corresponding to charmed hadrons does not show any significant change in both formulations with respect to no fluctuations scenario. We also found that the mass hierarchy of the kinetic freeze out parameters as argued in our previous analysis is still preserved even in presence of transverse flow fluctuations.

To the best of our knowledge, this is the first attempt to incorporate the flow fluctuations into the non-boost-invariant blast-wave model to describe the transverse momentum spectra of identified hadrons at SPS energies. As mentioned earlier, authors of Ref.~\cite{Akkelin:2009nx} have implemented the flow fluctuations into the boost-invariant blast-wave model to study the heavy hadrons namely, $J/\psi$, $\phi$ and $\Omega$. There was an attempt made to consider the transverse flow fluctuations in non-central collisions by the authors of Refs.~\cite{Voloshin:2008dg,Ollitrault:2009ie}. The authors have also used Bessel-Gaussian formulations for the descriptions of the initial state eccentricity fluctuations which are not purely Gaussian especially for peripheral collisions. Since in this article, we are exclusively dealing with central collisions, we refrain from using the Bessel Gaussian formulation.

The organization of the article is as follows: Following the introduction in this section, the features of the blast-wave model and its modification for incorporating transverse flow fluctuations is described in section II. The results are presented and discussed in section III. In section IV we summarize and conclude our findings from this study.

\section{A brief description of the model}

\begin{figure*}[t]
\begin{picture}(160,140)
\put(0,0){\includegraphics[scale=0.28]{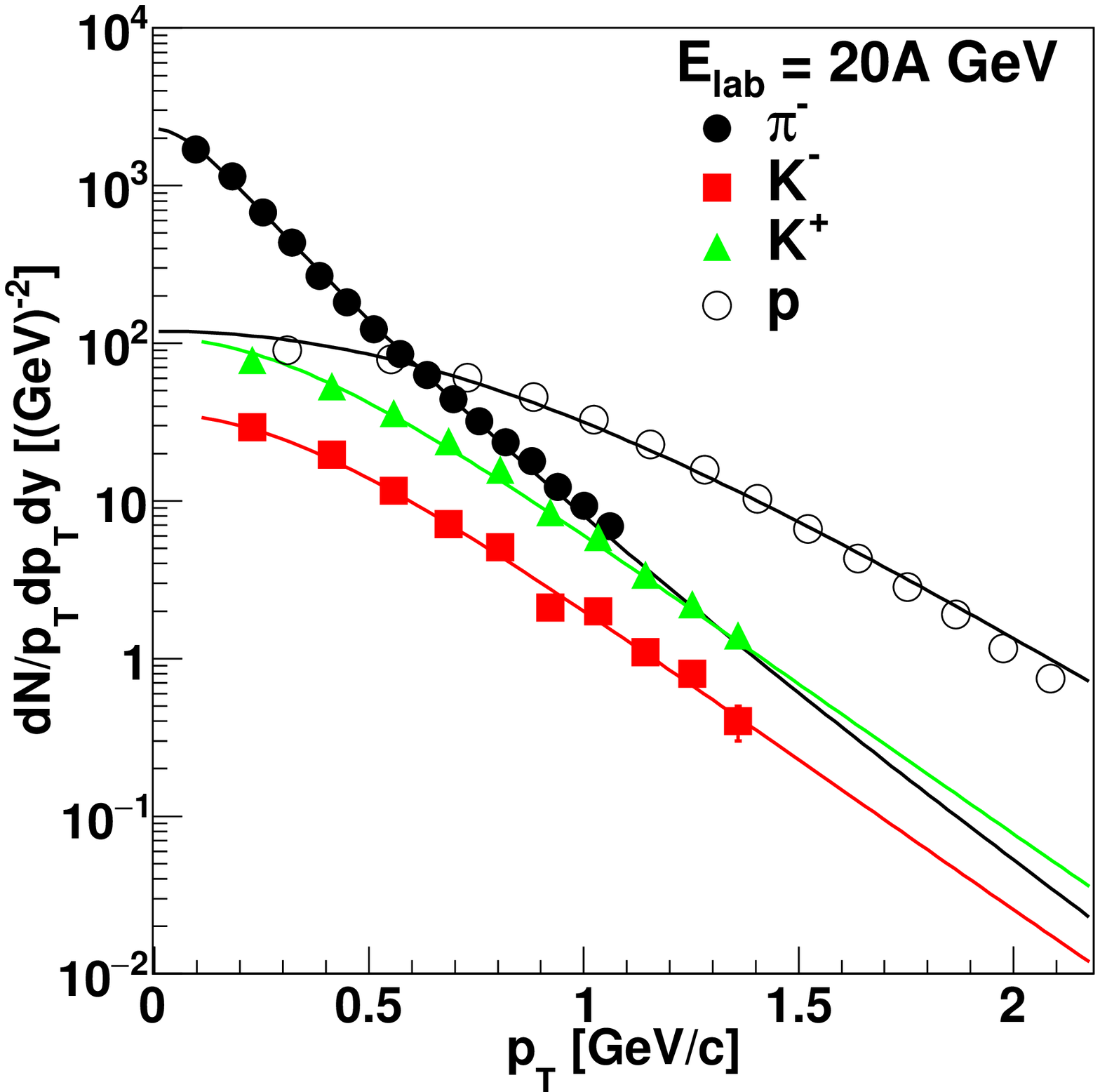}}
\put(50,125){(a)}
\end{picture}
\begin{picture}(160,140)
\put(0,0){\includegraphics[scale=0.28]{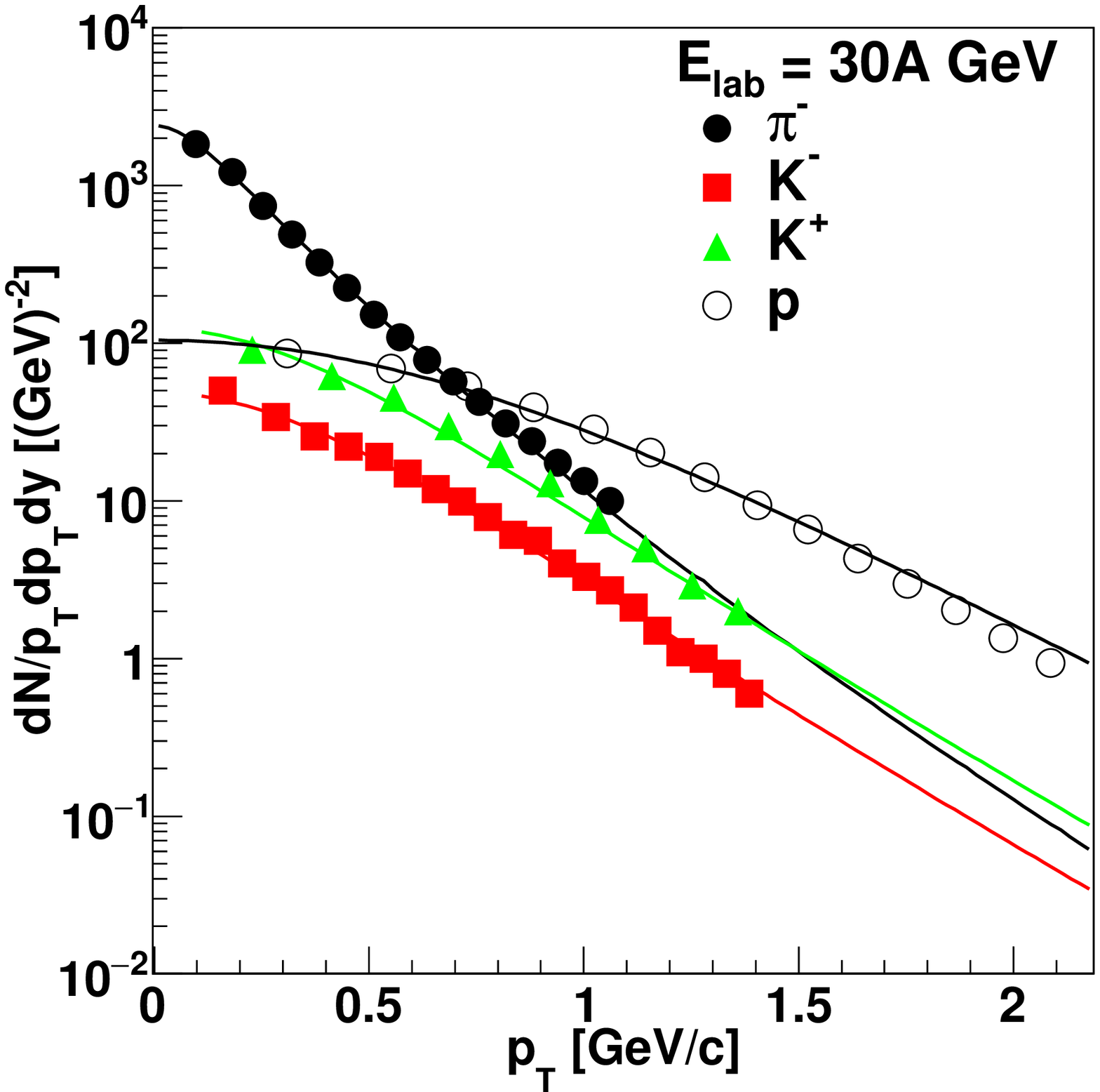}}
\put(50,125){(b)}
\end{picture}
\begin{picture}(160,160)
\put(0,0){\includegraphics[scale=0.28]{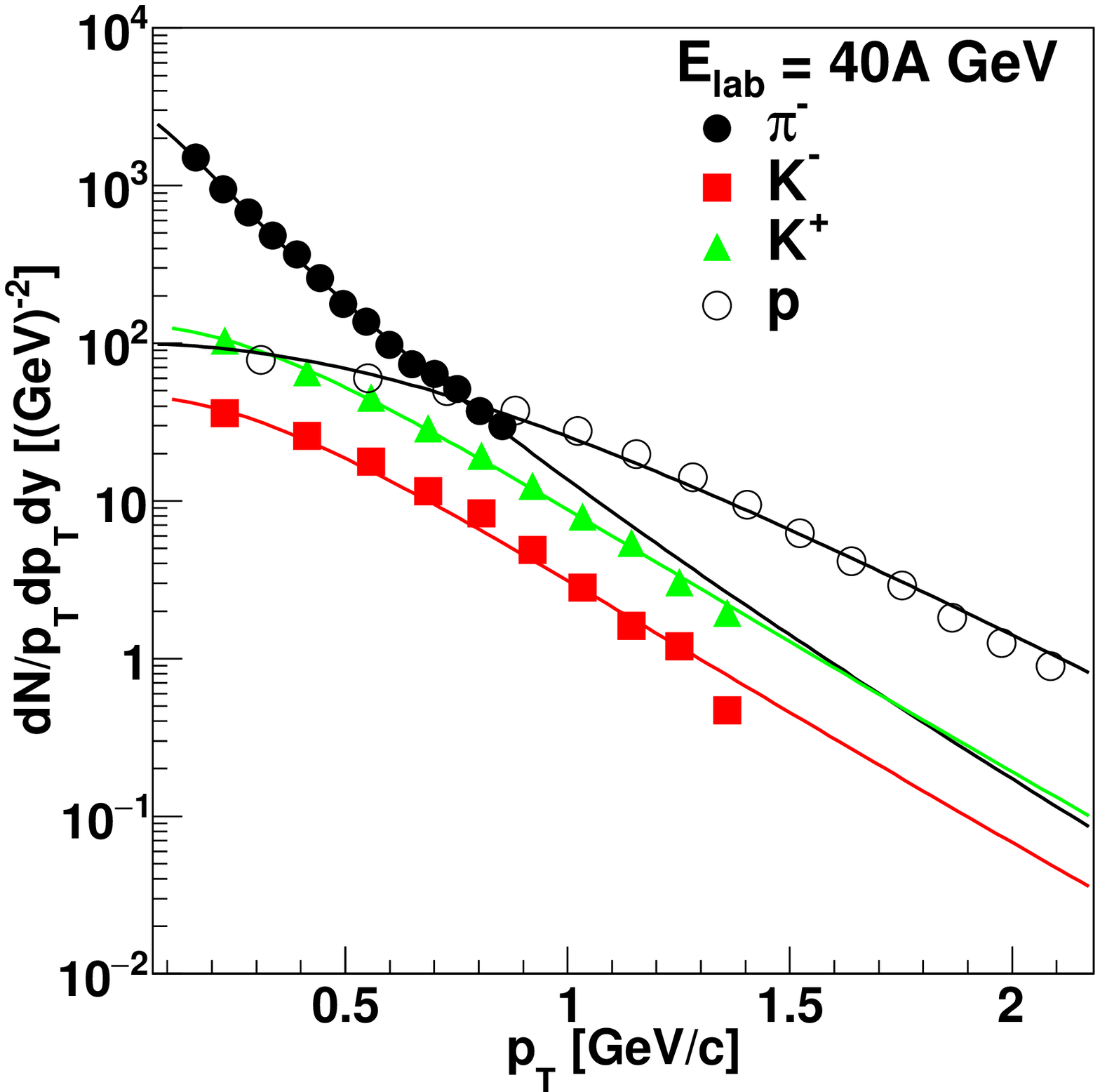}}
\put(50,125){(c)}
\end{picture}
\begin{picture}(160,160)
\put(0,0){\includegraphics[scale=0.28]{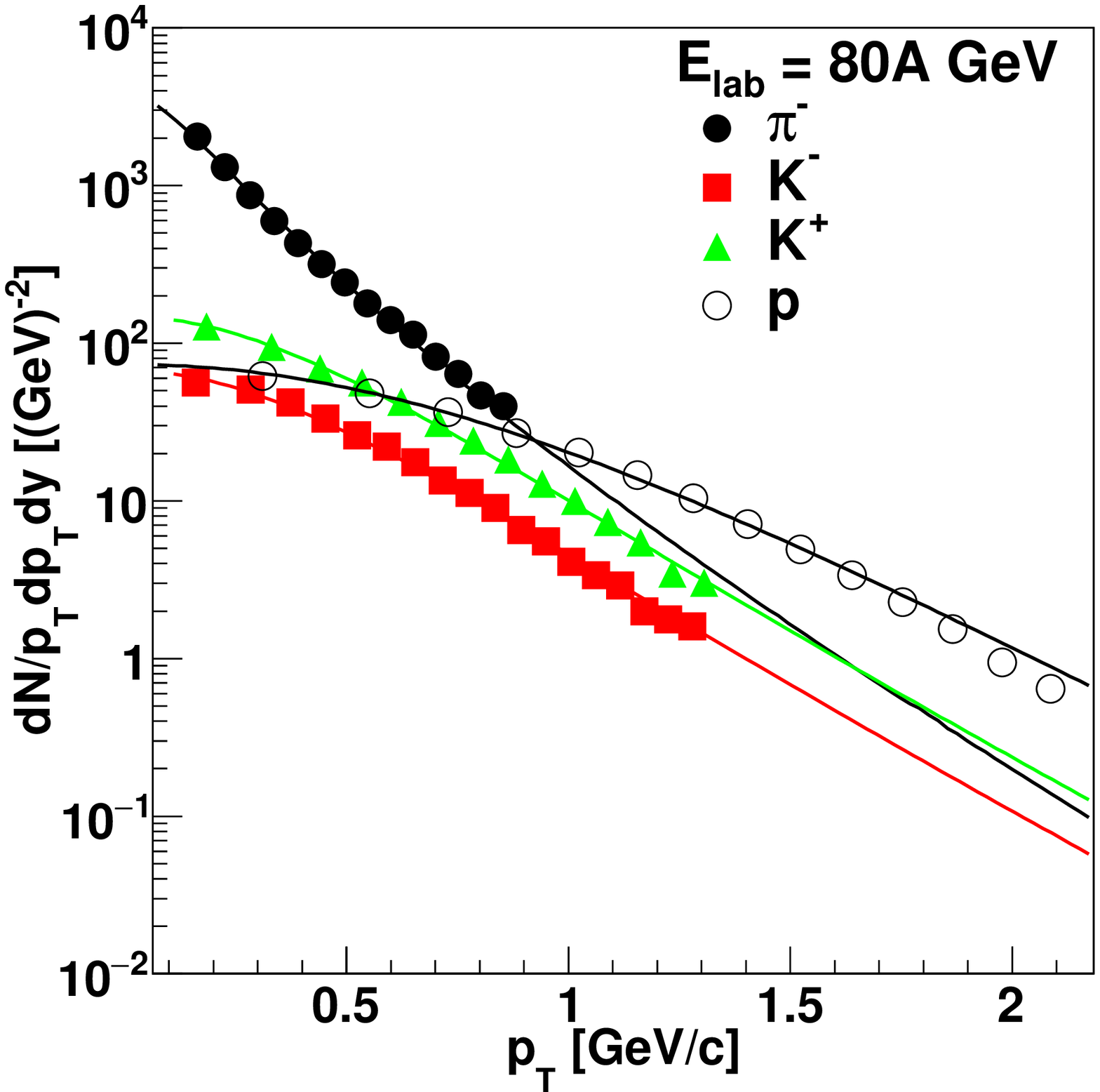}}
\put(50,125){(d)}
\end{picture}
\begin{picture}(160,160)
\put(0,0){\includegraphics[scale=0.28]{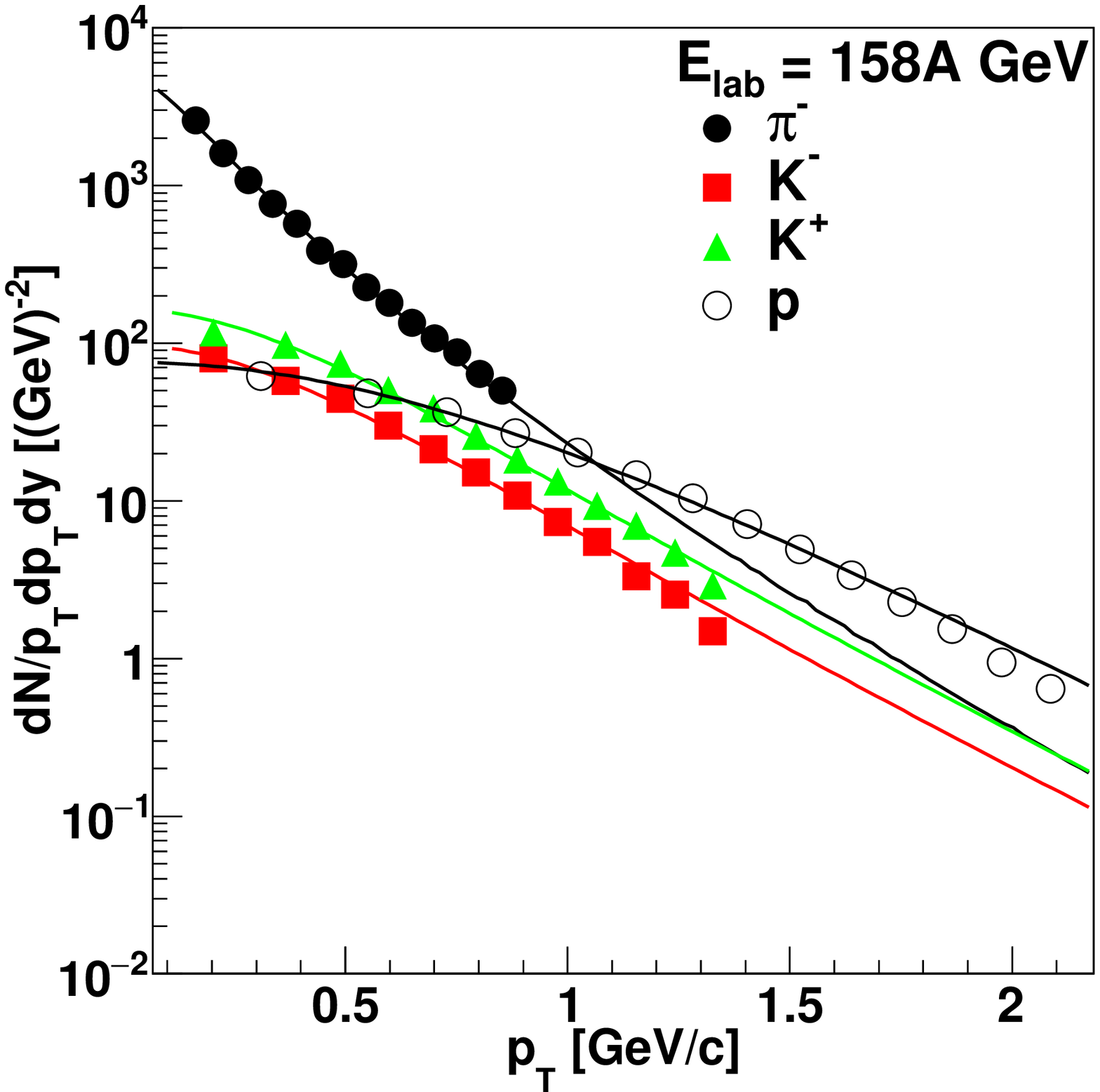}}
\put(50,125){(e)}
\end{picture}
\caption{Simultaneously fitted $p_{T}$-spectra of $\pi^{-}$, $\rm K^{\pm}$, and $\rm p$ at (a) 20A GeV, (b) 30A GeV, (c) 40A GeV, (d) 80A GeV and  (e) 158A GeV beam energies using uniform profile of transverse flow fluctuations. Error bars indicate available statistical error.}
\label{figl1}
\end{figure*}

\begin{figure*}[h]
\begin{picture}(160,140)
\put(0,0){\includegraphics[scale=0.28]{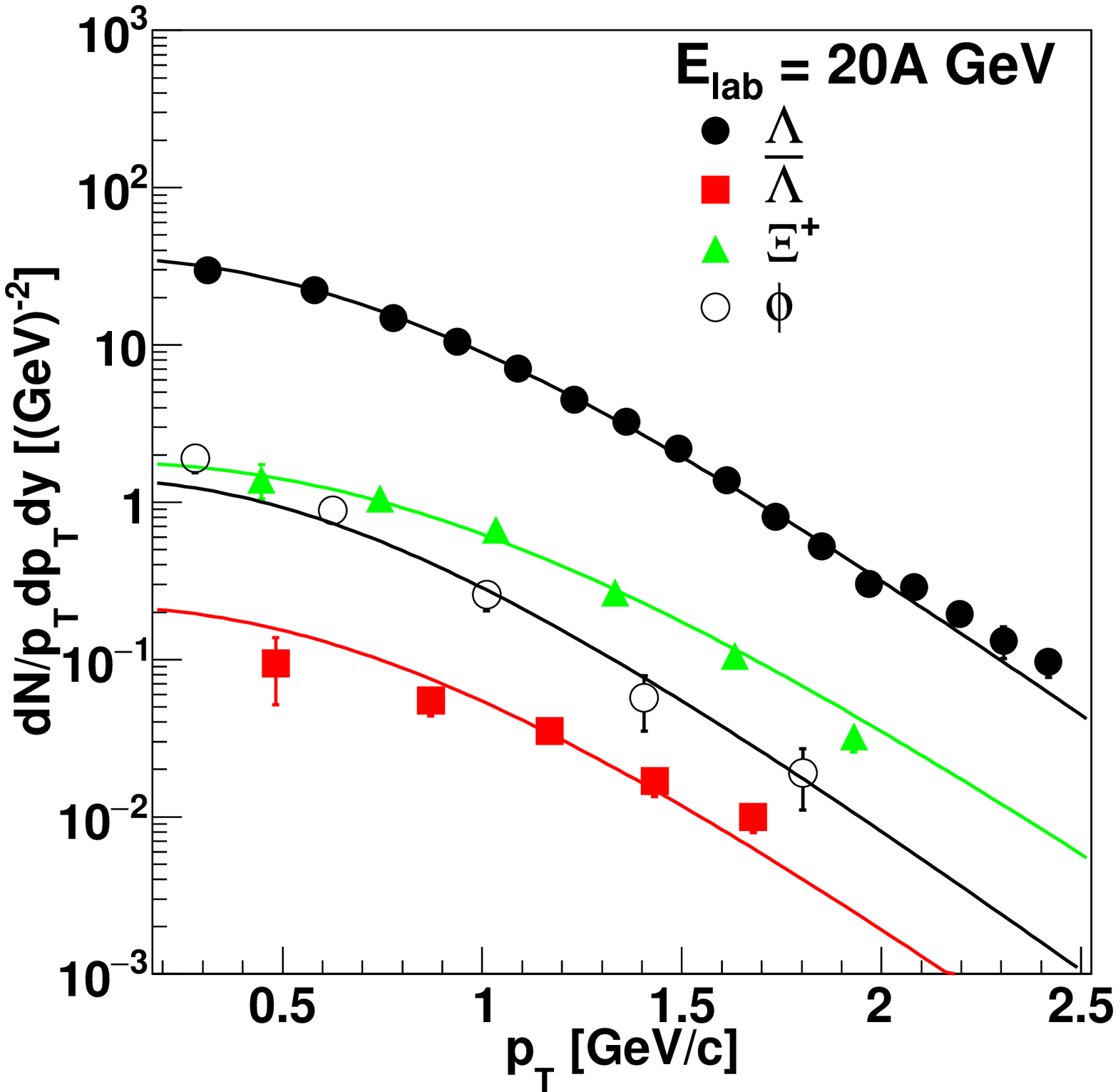}}
\put(50,125){(a)}
\end{picture}
\begin{picture}(160,140)
\put(0,0){\includegraphics[scale=0.28]{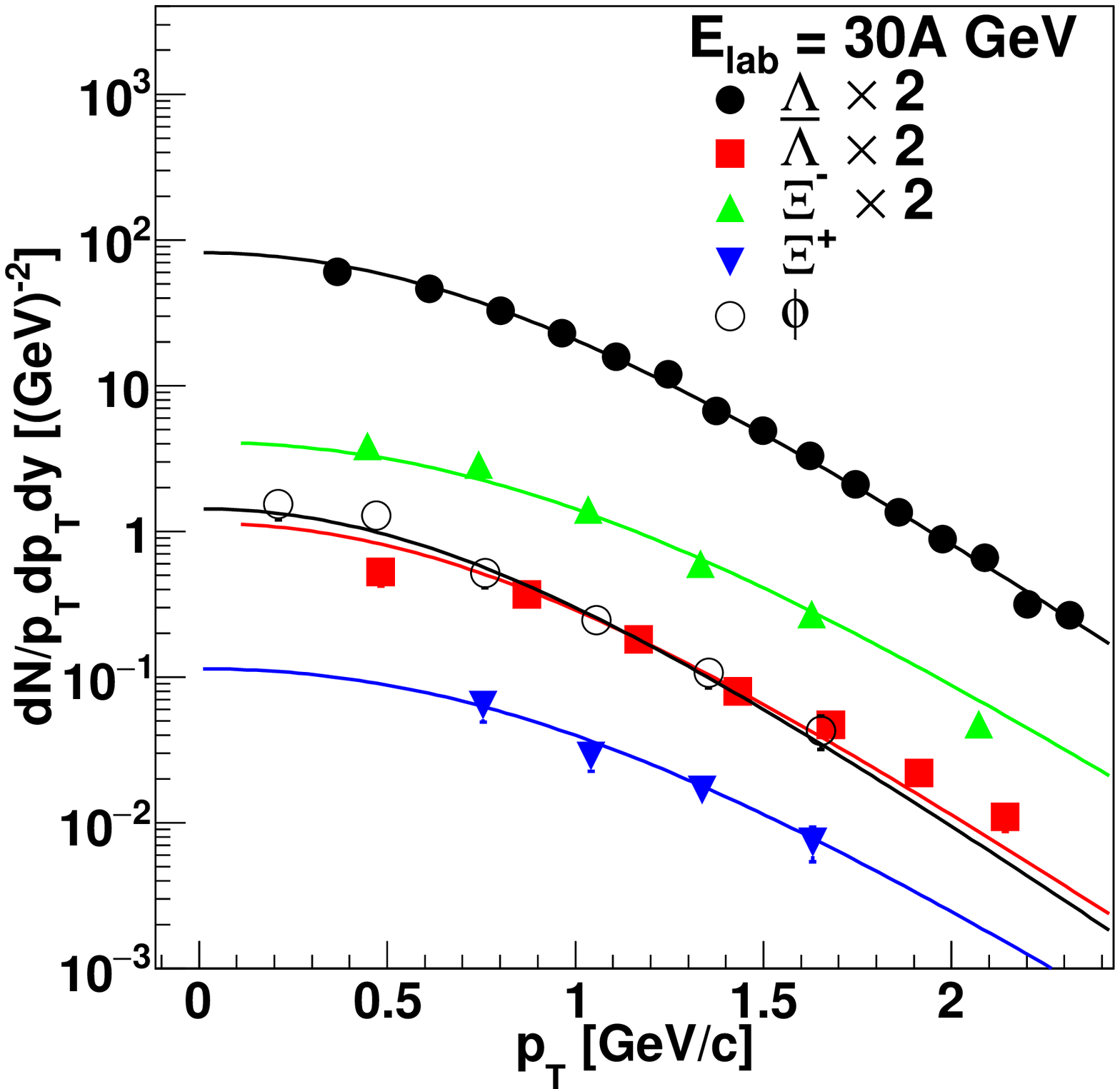}}
\put(50,125){(b)}
\end{picture}
\begin{picture}(160,160)
\put(0,0){\includegraphics[scale=0.28]{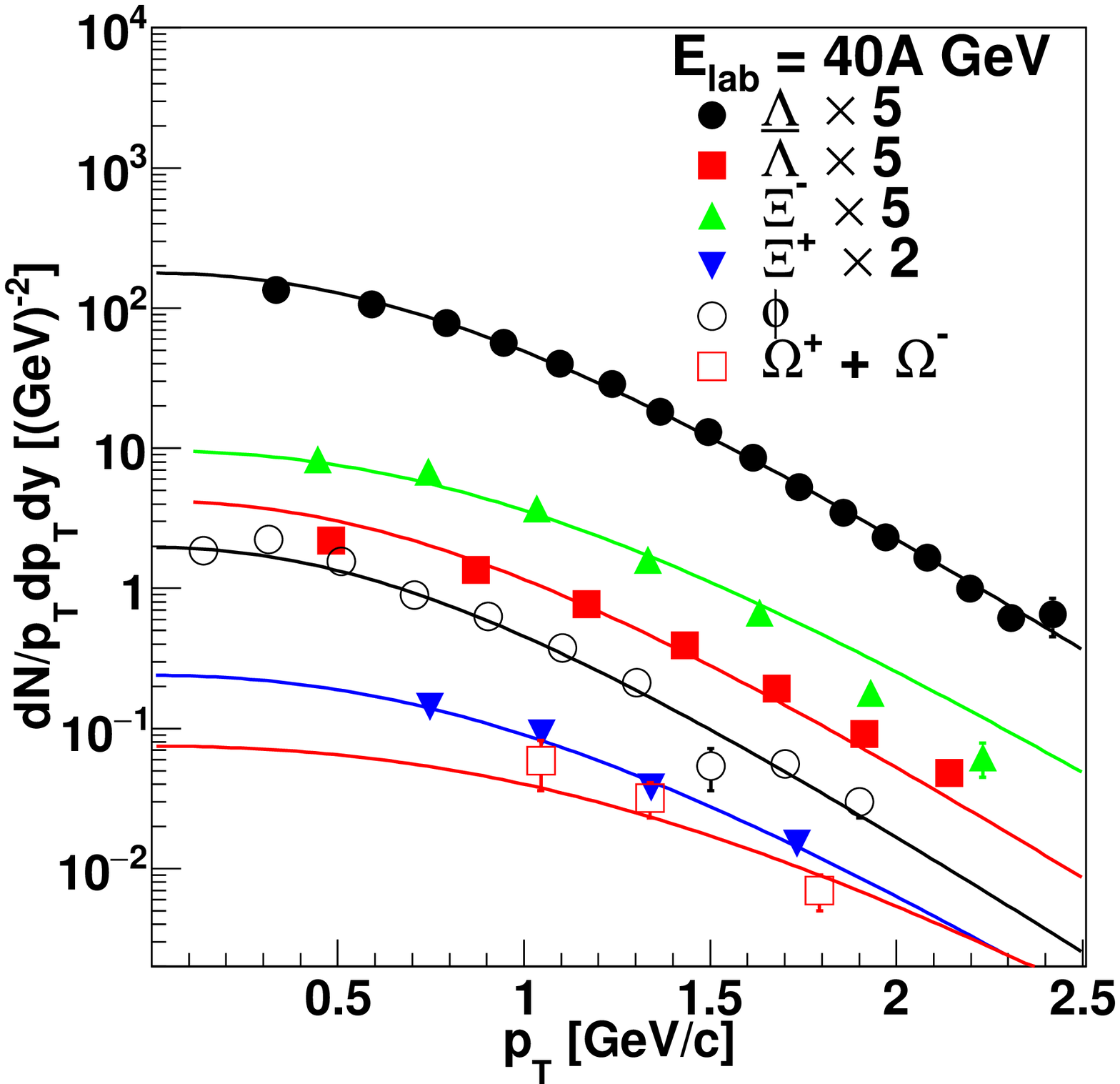}}
\put(50,125){(c)}
\end{picture}
\begin{picture}(160,160)
\put(0,0){\includegraphics[scale=0.28]{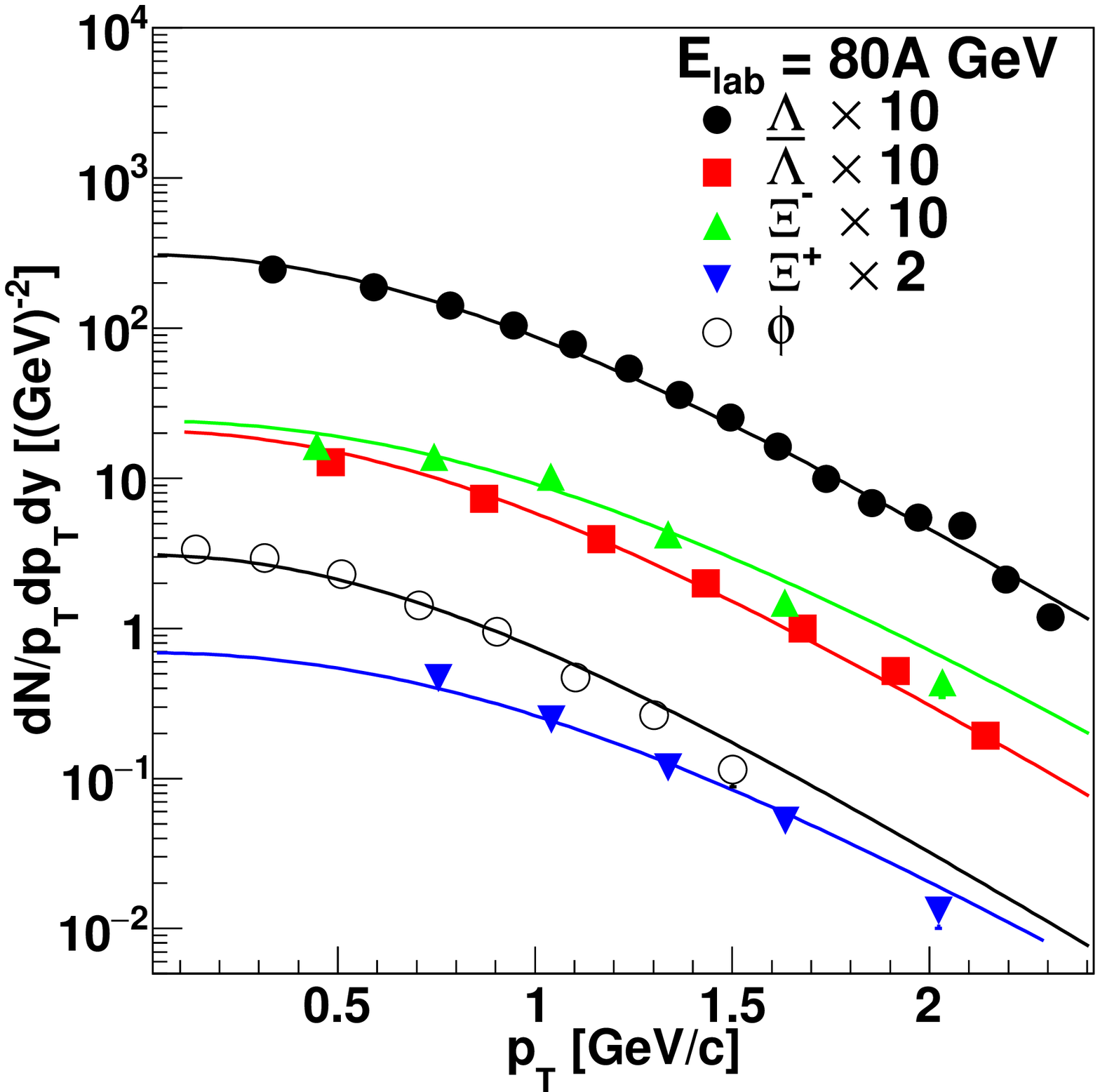}}
\put(50,125){(d)}
\end{picture}
\begin{picture}(160,160)
\put(0,0){\includegraphics[scale=0.28]{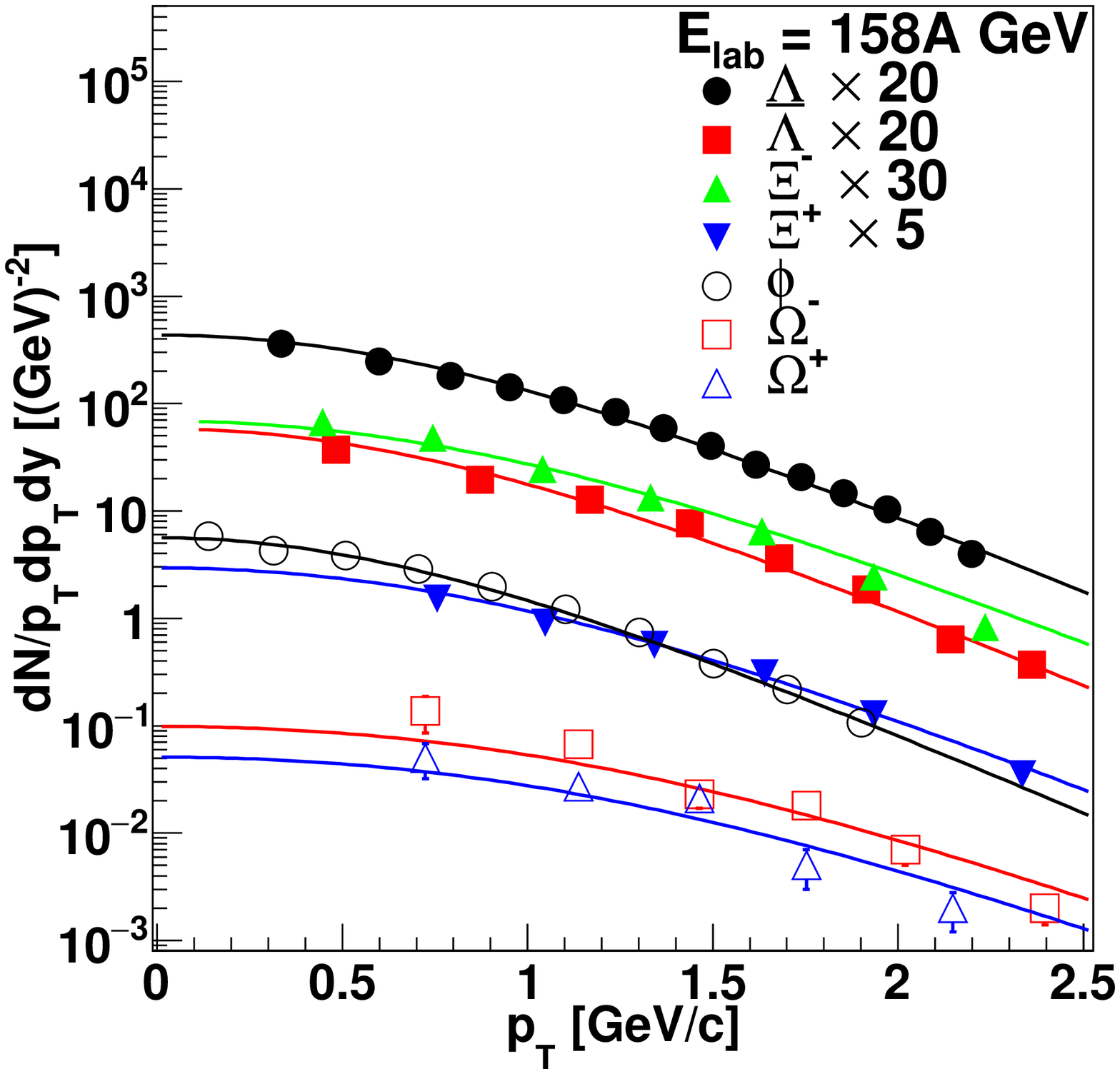}}
\put(50,125){(e)}
\end{picture}
\caption{Simultaneously fitted $p_{T}$-spectra of $\Lambda$, $\bar{\Lambda}$, $\phi$, $\Xi^{\pm}$ and $\Omega^{\pm}$ at (a) 20A GeV, (b) 30A GeV, (c) 40A GeV, (d) 80A GeV and  (e) 158A GeV beam energies using uniform profile of transverse flow fluctuations. Error bars indicate available statistical error.}
\label{fig1}
\end{figure*}

\begin{table*}[h]\centering
\vglue4mm
\begin{tabular}{|c|c|c|c|c|c|c|c|c|c|} \hline
Species & $\rm E_{Lab}$ & $\eta_{\rm max}$ & $\beta^{min}_{s}$   &$\beta^{max}_{s}$  & $\beta^{0}_{s}$  & $T_{kin}$& $\chi^{2}/N_{\rm dof}$\\
& (A GeV) &  &   &  &   & (MeV) & \\ \hline

$\pi^{-}$, $\rm K^{\pm}$, $\rm p$& 20 & $1.882  \pm  0.005$ & $0.653  \pm  0.002$ & $0.852  \pm  0.003$ & $0.752 \pm 0.004$ ($0.777  \pm  0.002$) & $91.62  \pm  0.22$ ($79.78  \pm  0.05$) & 6.7 (6.5)\\
& 30 & $2.084  \pm  0.004$ & $0.618  \pm  0.003$ & $0.926  \pm  0.004$ & $0.772 \pm 0.005$ ($0.805  \pm  0.002$) & $93.51  \pm  0.23$ ($80.28  \pm  0.05$) & 7.2 (6.7)\\
& 40 & $2.094  \pm  0.004$ & $0.596  \pm  0.003$ & $0.873  \pm  0.005$ & $0.734 \pm 0.005$ ($0.803  \pm  0.001$) & $108.97  \pm  0.38$ ($81.92  \pm  0.04$) & 5.6 (5.5)\\
& 80 & $2.391  \pm  0.005$ & $0.631  \pm  0.003$ & $0.914  \pm  0.006$ & $0.772 \pm 0.007$ ($0.802  \pm  0.002$) & $97.40  \pm  0.40$ ($82.68  \pm  0.05$) & 3.7 (3.8)\\
& 158 & $2.621  \pm  0.006$ & $0.601  \pm  0.004$ & $0.925  \pm  0.006$ & $0.764 \pm 0.007$ ($0.807  \pm  0.002$) & $104.41  \pm  0.44$ ($84.11  \pm  0.05$) & 4.5 (4.4)\\

\hline
\hline

$\Lambda$, $\bar{\Lambda}$, $\phi$,& 20 & $1.288  \pm  0.021$ & $0.515  \pm  0.021$ & $0.744  \pm  0.023$ & $0.630 \pm 0.016$ ($0.663  \pm  0.005$) & $105.17  \pm  1.53$ ($93.12  \pm  0.19$) & 1.5 (1.8)\\
$\Xi^{\pm}$, $\Omega^{\pm}$& 30 & $1.728  \pm  0.026$ & $0.507  \pm  0.021$ & $0.772  \pm  0.016$ & $0.639 \pm 0.013$ ($0.675  \pm  0.004$) & $105.50  \pm  1.06$ ($95.84  \pm  0.17$) & 1.9 (2.2)\\
& 40 & $1.752  \pm  0.018$ & $0.541  \pm  0.014$ & $0.762  \pm  0.016$ & $0.652 \pm 0.011$ ($0.681  \pm  0.004$) & $110.46  \pm  1.17$ ($98.87  \pm  0.13$) & 3.6 (3.6)\\
& 80 & $1.989  \pm  0.021$ & $0.554  \pm  0.008$ & $0.722  \pm  0.014$ & $0.638 \pm 0.008$ ($0.673  \pm  0.003$) & $124.51  \pm  1.48$ ($106.54  \pm  0.12$) & 3.5 (3.4)\\
& 158 & $2.031  \pm  0.029$ & $0.555  \pm  0.007$ & $0.733  \pm  0.011$ & $0.644 \pm 0.006$ ($0.703  \pm  0.002$) & $135.99  \pm  1.24$ ($109.24  \pm  0.11$) & 3.4 (3.4)\\

\hline
\end{tabular}
\caption{Summary of the fit results of $p_{\rm T}$-spectra of light and heavy strange hadrons after implementing the flow fluctuations with uniform distribution of transverse velocity, at different energies ranging from 20A to 158A GeV at SPS. The values $\eta_{max}$ are kept the same as no fluctuations scenario and adopted from Refs~\cite{Rode:2020vhu} and~\cite{Rode:2018hlj}. The corresponding fit results in no fluctuations scenario are quoted in parenthesis.}
\label{tabII}
\end{table*}

\begin{figure*}[t]
\begin{picture}(160,140)
\put(0,0){\includegraphics[scale=0.28]{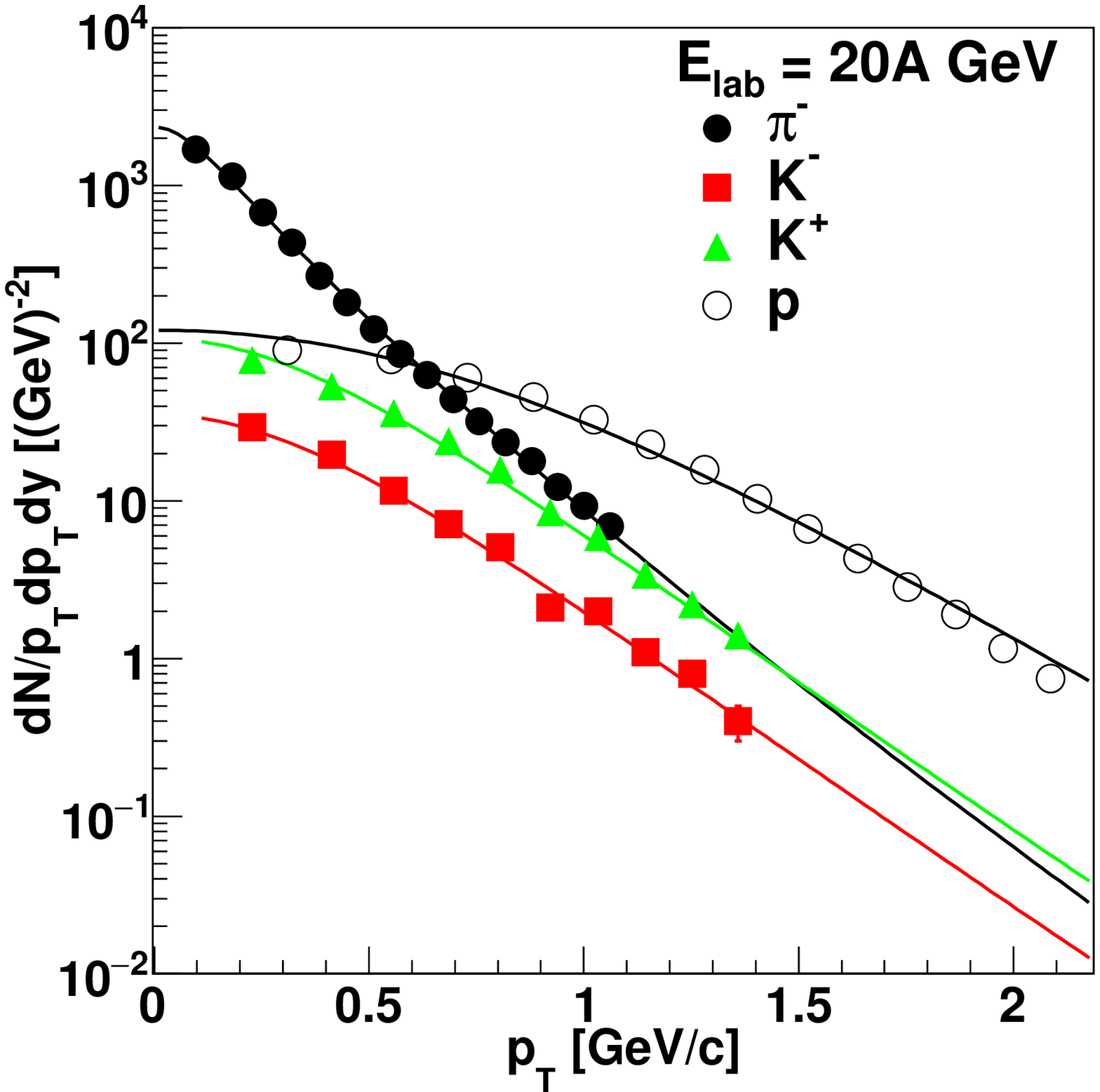}}
\put(50,125){(a)}
\end{picture}
\begin{picture}(160,140)
\put(0,0){\includegraphics[scale=0.28]{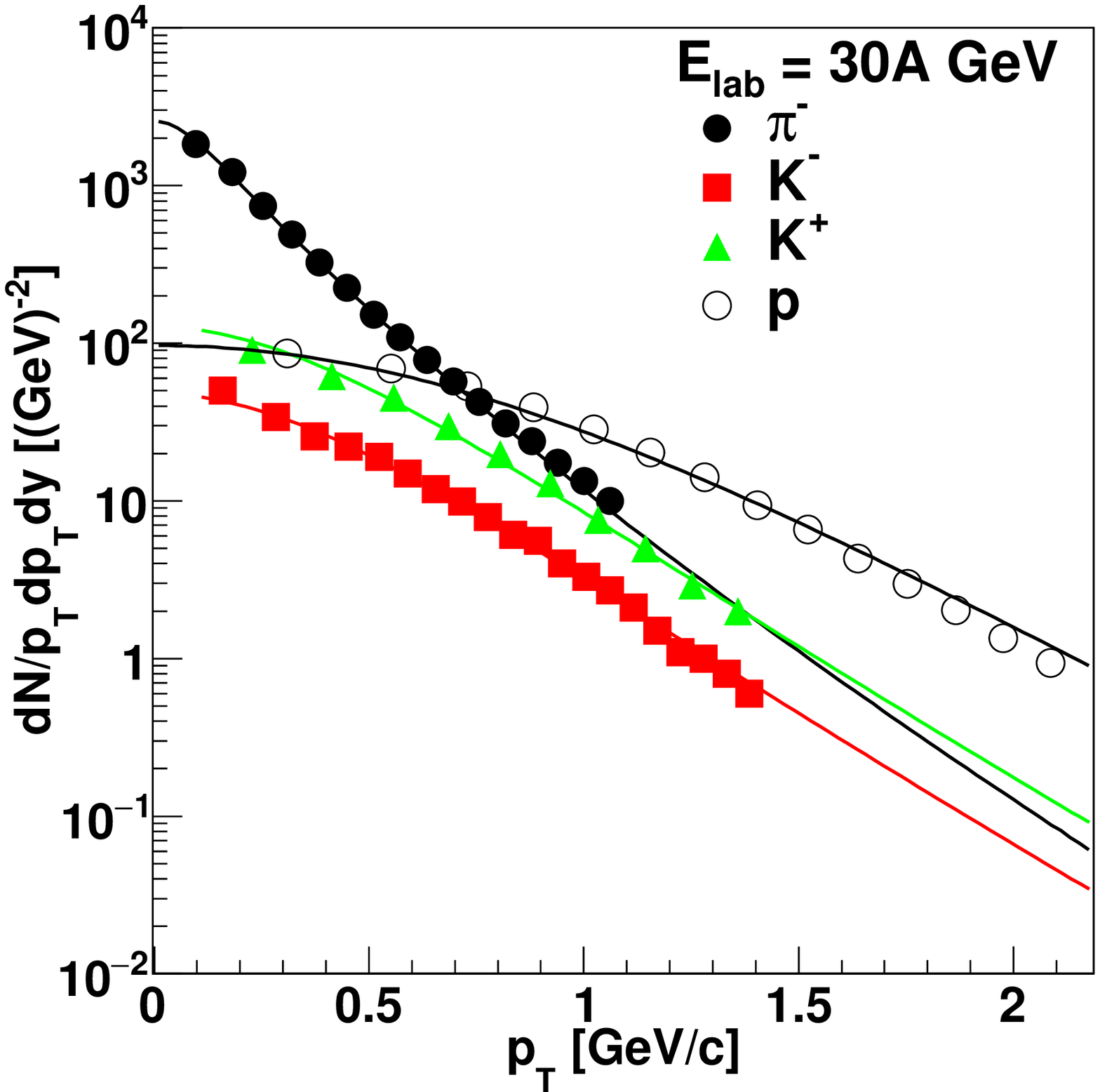}}
\put(50,125){(b)}
\end{picture}
\begin{picture}(160,160)
\put(0,0){\includegraphics[scale=0.28]{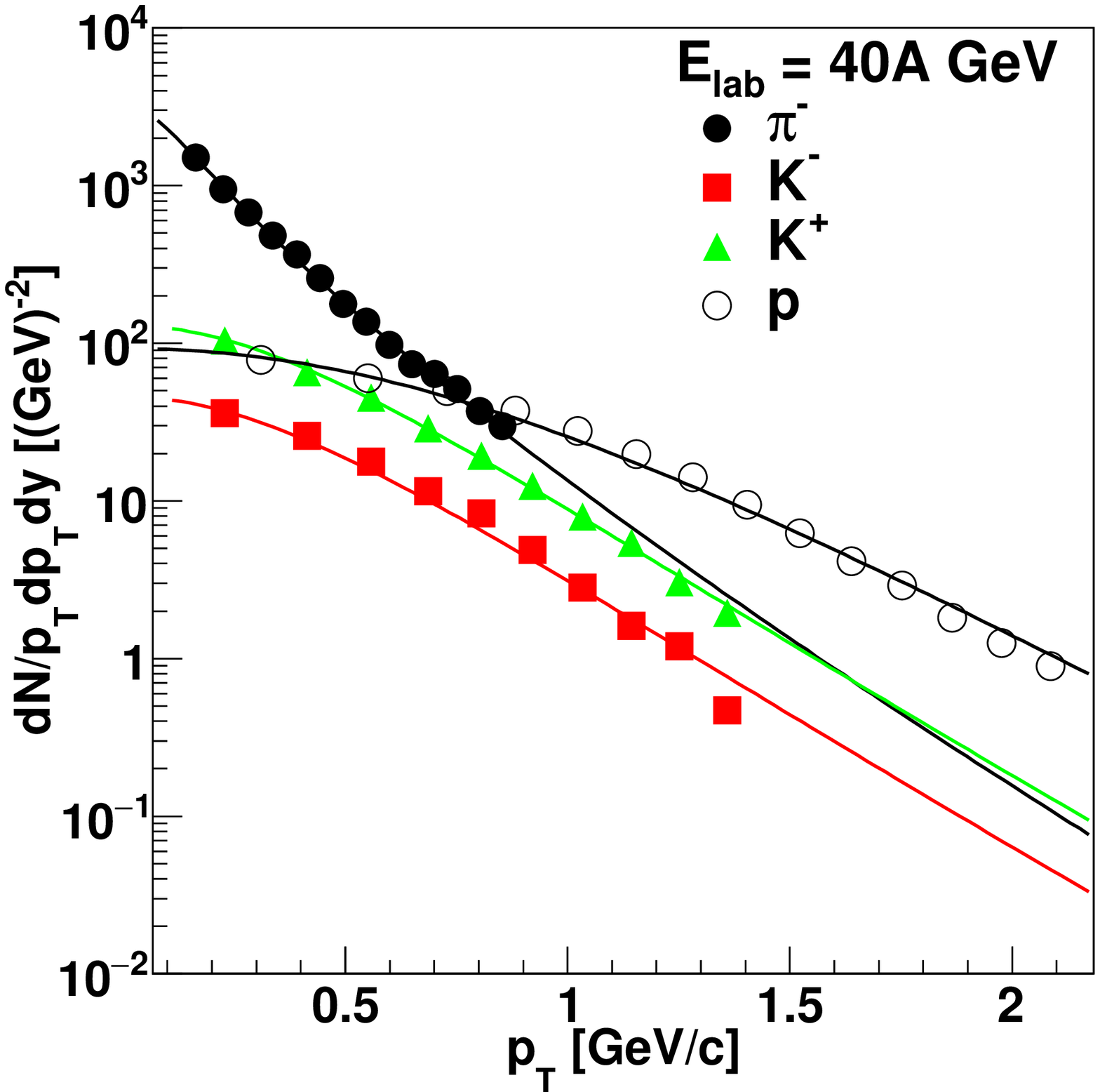}}
\put(50,125){(c)}
\end{picture}
\begin{picture}(160,160)
\put(0,0){\includegraphics[scale=0.28]{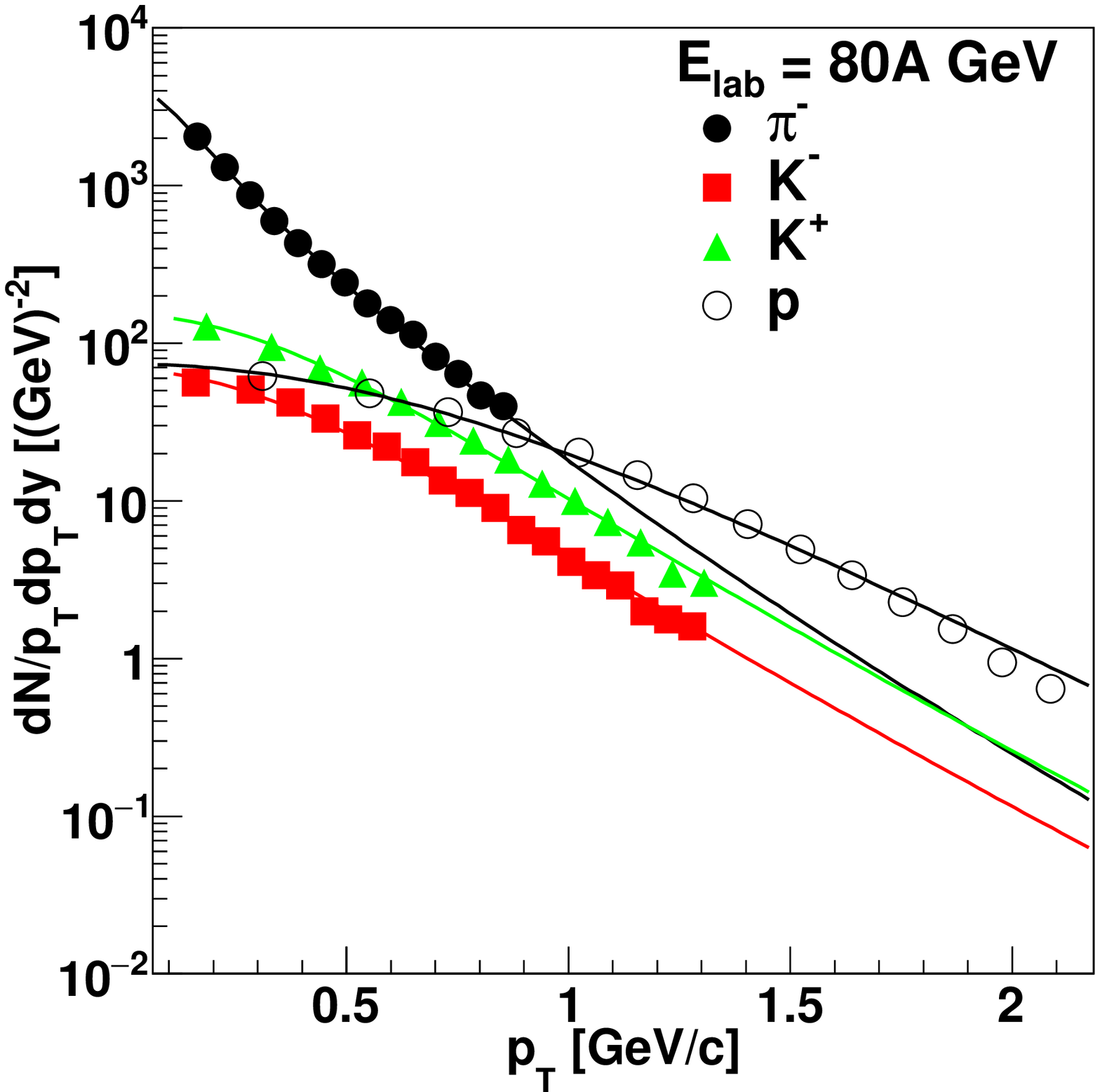}}
\put(50,125){(d)}
\end{picture}
\begin{picture}(160,160)
\put(0,0){\includegraphics[scale=0.28]{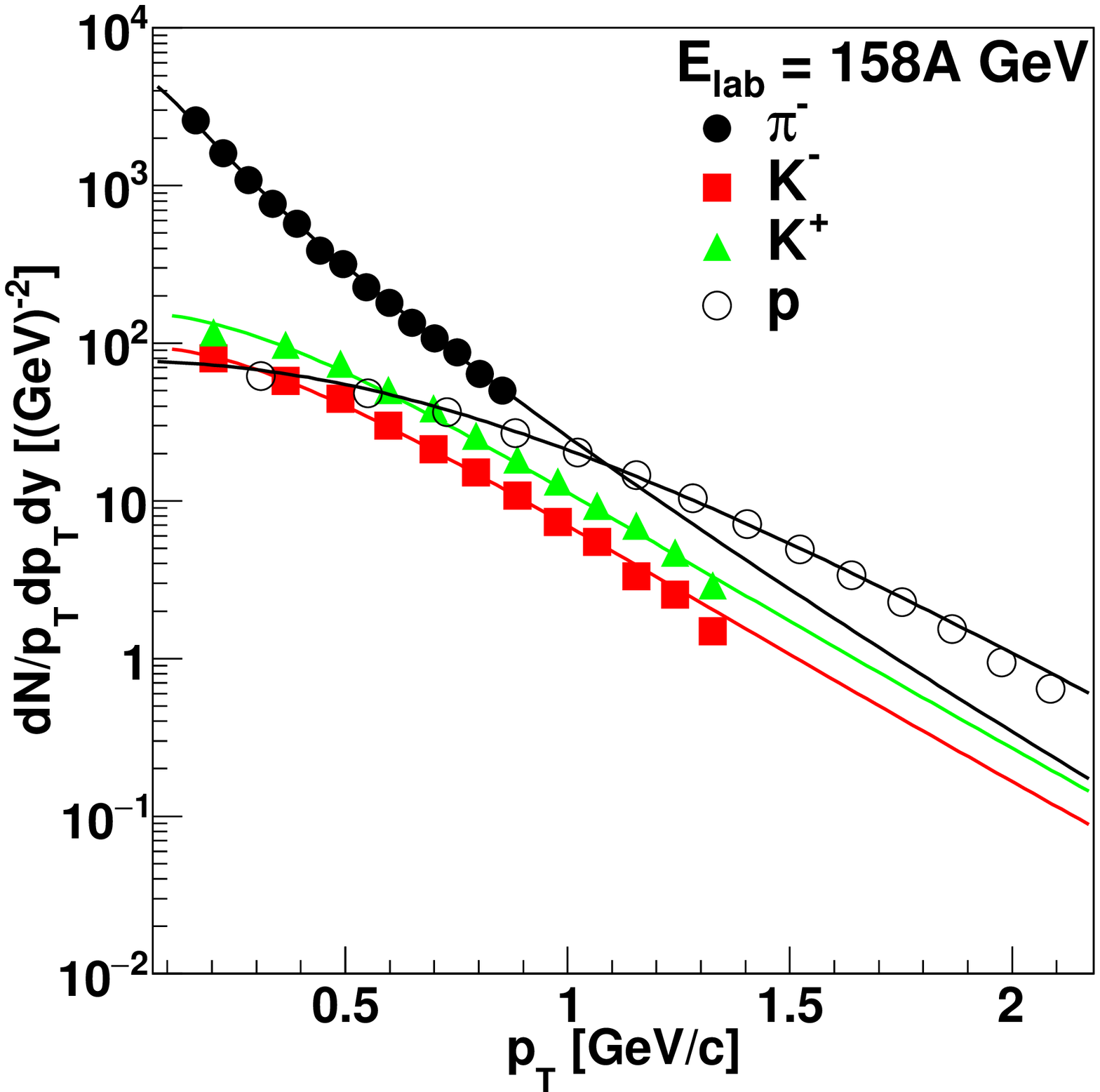}}
\put(50,125){(e)}
\end{picture}
\caption{Simultaneously fitted $p_{T}$-spectra of $\pi^{-}$, $\rm K^{\pm}$, and $\rm p$ at (a) 20A GeV, (b) 30A GeV, (c) 40A GeV, (d) 80A GeV and  (e) 158A GeV beam energies using Gaussian description of transverse flow fluctuations. Error bars indicate available statistical error.}
\label{figs1}
\end{figure*}

\begin{figure*}[h]
\begin{picture}(160,140)
\put(0,0){\includegraphics[scale=0.28]{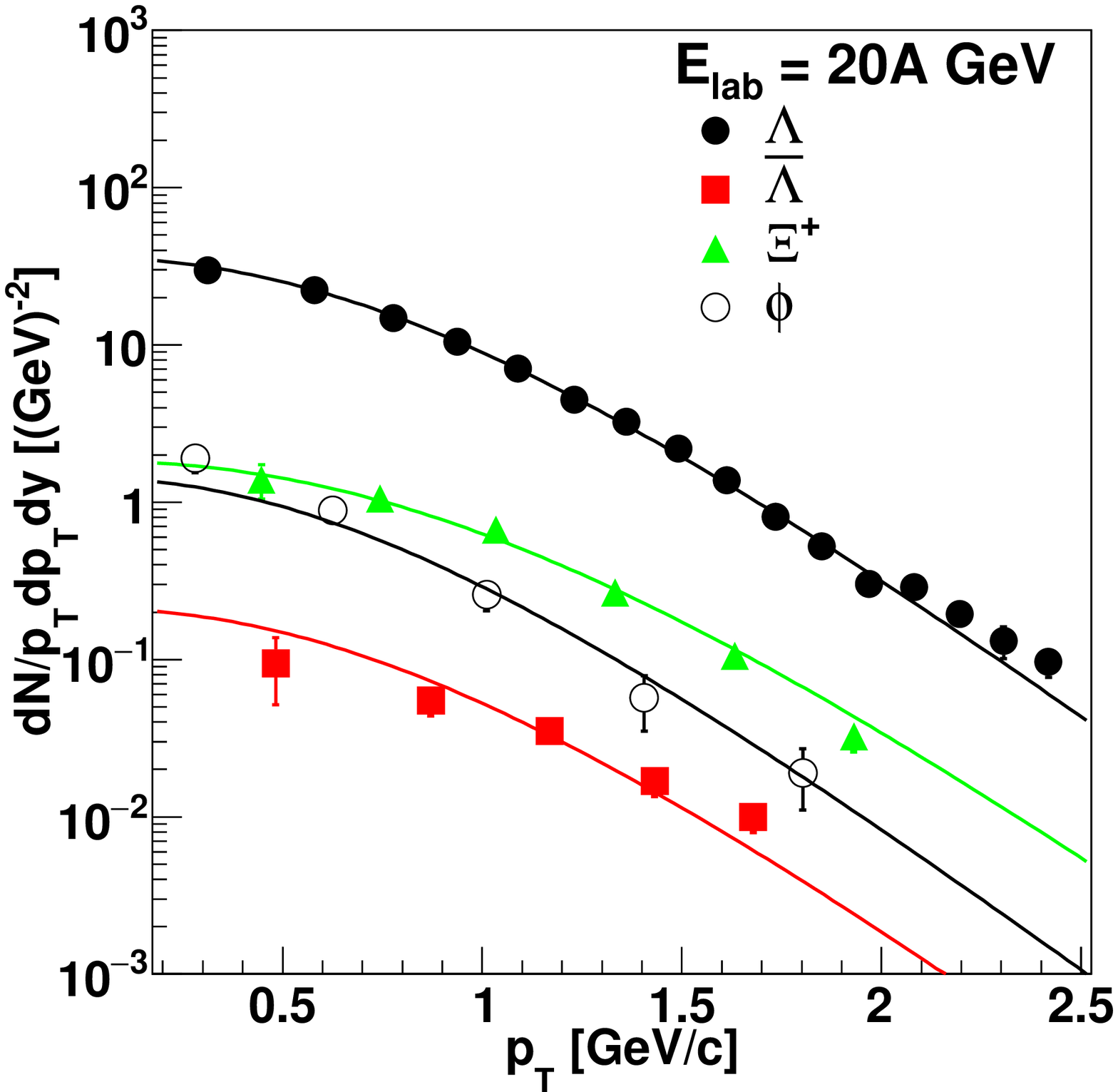}}
\put(50,125){(a)}
\end{picture}
\begin{picture}(160,140)
\put(0,0){\includegraphics[scale=0.28]{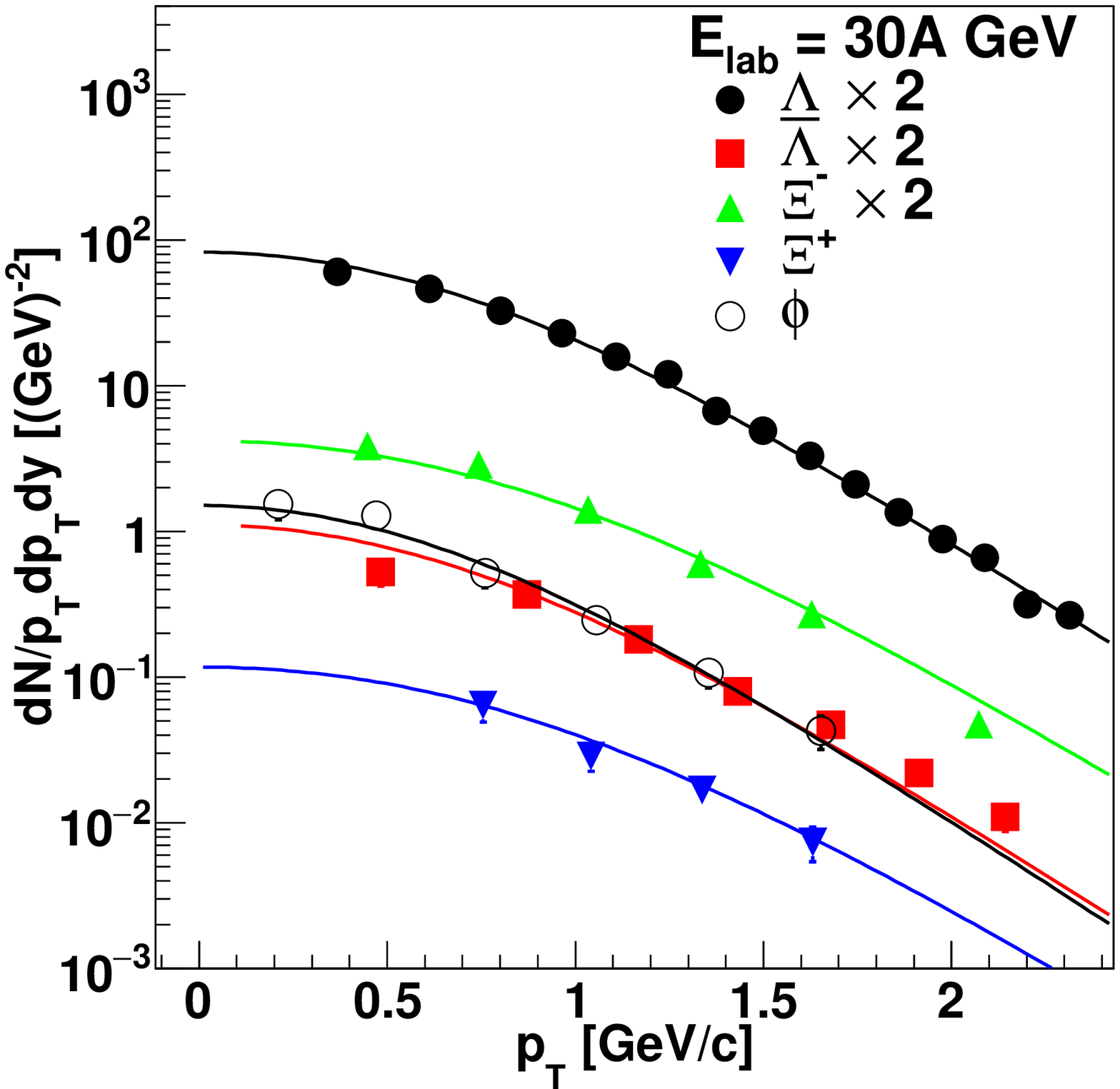}}
\put(50,125){(b)}
\end{picture}
\begin{picture}(160,160)
\put(0,0){\includegraphics[scale=0.28]{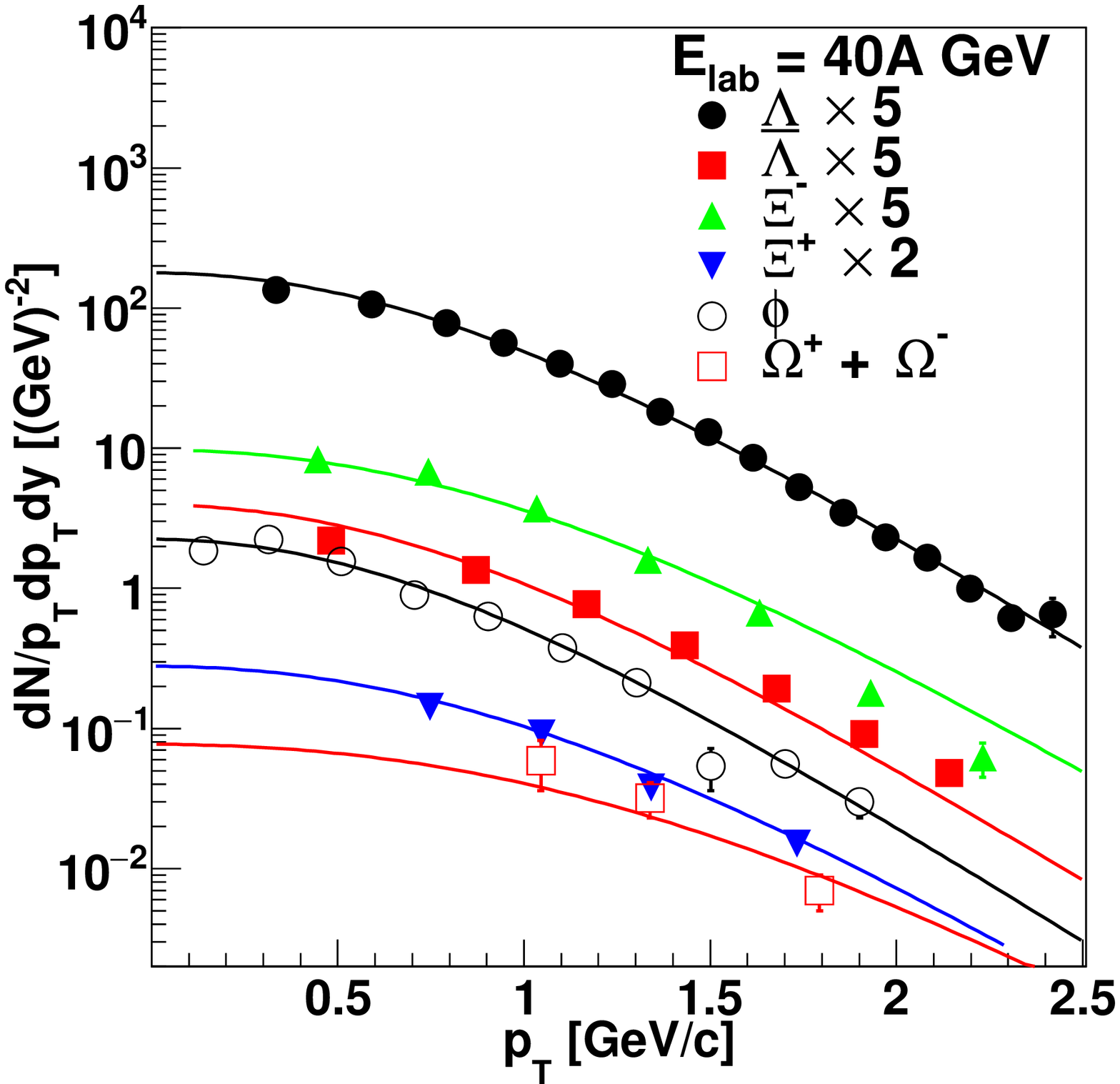}}
\put(50,125){(c)}
\end{picture}
\begin{picture}(160,160)
\put(0,0){\includegraphics[scale=0.28]{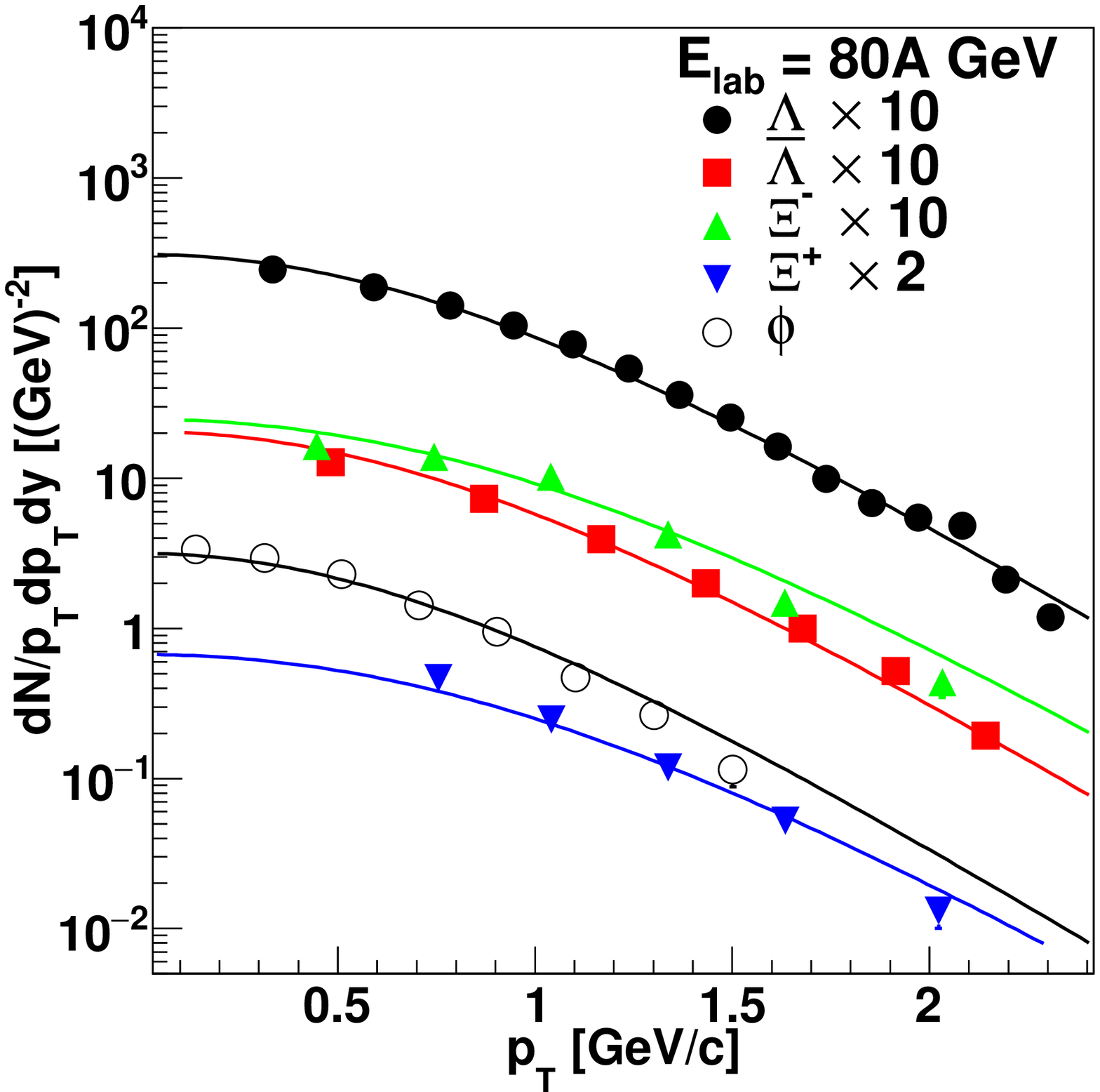}}
\put(50,125){(d)}
\end{picture}
\begin{picture}(160,160)
\put(0,0){\includegraphics[scale=0.28]{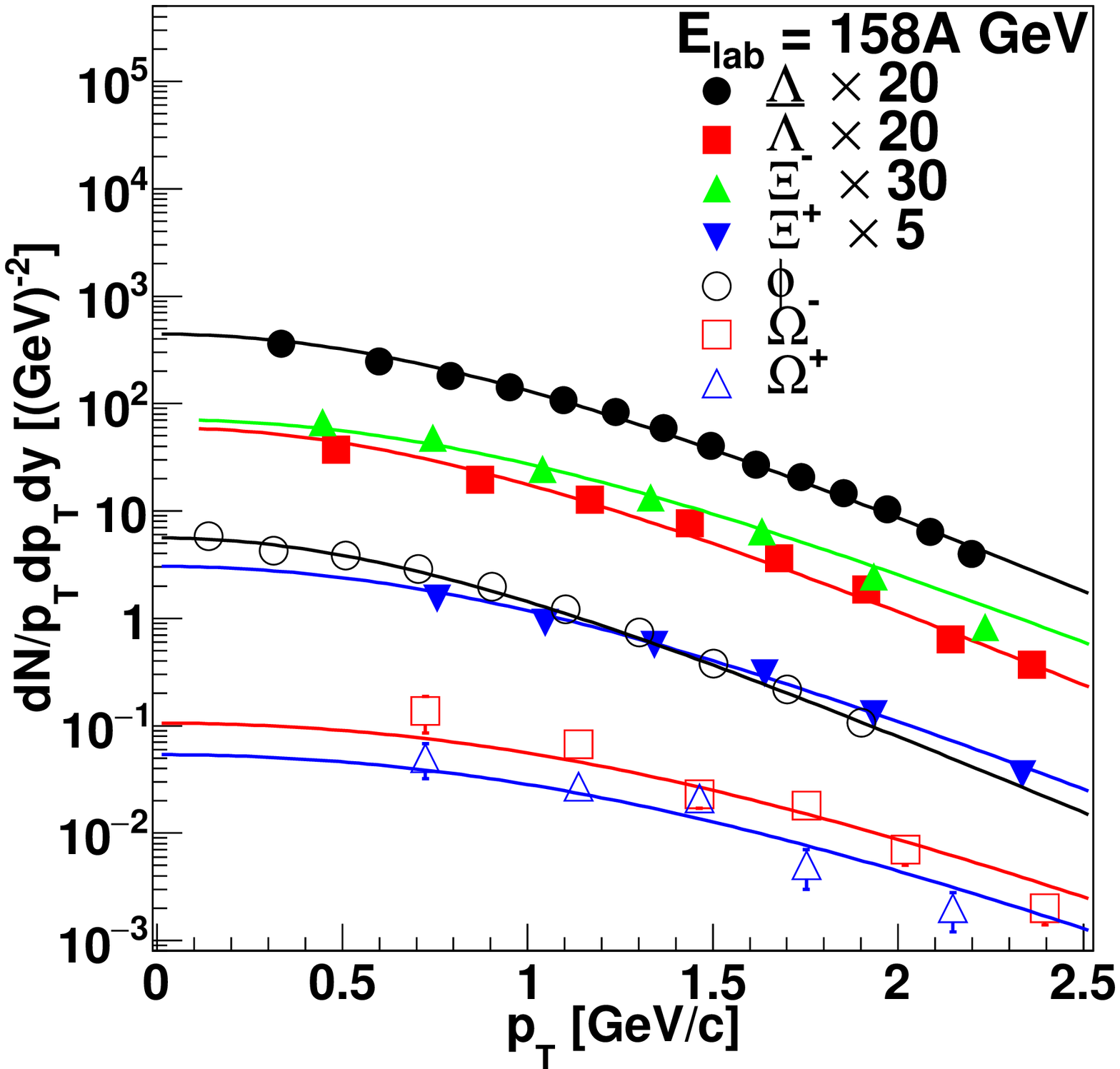}}
\put(50,125){(e)}
\end{picture}
\caption{Simultaneously fitted $p_{T}$-spectra of $\Lambda$, $\bar{\Lambda}$, $\phi$, $\Xi^{\pm}$ and $\Omega^{\pm}$ at (a) 20A GeV, (b) 30A GeV, (c) 40A GeV, (d) 80A GeV and  (e) 158A GeV beam energies using Gaussian description of transverse flow fluctuations. Error bars indicate available statistical error.}
\label{fig3}
\end{figure*}

\begin{table*}[h]\centering
\vglue4mm
\begin{tabular}{|c|c|c|c|c|c|c|c|} \hline
Species & $\rm E_{Lab}$~(A GeV) & $\eta_{\rm max}$ & $\beta^{0}_{s}$  &$\delta$ & $T_{kin}$ (MeV) & $\chi^{2}/N_{\rm dof}$\\ \hline

$\pi^{-}$, $\rm K^{\pm}$, $\rm p$ & 20 & $1.882  \pm  0.005$ & $0.736  \pm  0.002$ ($0.777  \pm  0.002$) & $0.085  \pm  0.001$ & $93.58  \pm  0.17$ ($79.78  \pm  0.05$) & 7.2 (6.5)\\
& 30 & $2.084  \pm  0.004$ & $0.767  \pm  0.002$ ($0.805  \pm  0.002$) & $0.109  \pm  0.002$ & $94.02  \pm  0.19$ ($80.28  \pm  0.05$) & 6.6 (6.7)\\
& 40 & $2.094  \pm  0.004$ & $0.744  \pm  0.002$ ($0.803  \pm  0.001$) & $0.095  \pm  0.002$ & $102.69  \pm  0.28$ ($81.92  \pm  0.04$) & 4.9 (5.5)\\
& 80 & $2.391  \pm  0.005$ & $0.747  \pm  0.003$ ($0.802  \pm  0.002$) & $0.127  \pm  0.003$ & $102.47  \pm  0.35$ ($82.68  \pm  0.05$) & 3.1 (3.8)\\
& 158 & $2.621  \pm  0.006$ & $0.738  \pm  0.003$ ($0.807  \pm  0.002$) & $0.084  \pm  0.002$ & $109.23  \pm  0.38$ ($84.11  \pm  0.05$) & 3.7 (4.4)\\

\hline
\hline

$\Lambda$, $\bar{\Lambda}$, $\phi$, $\Xi^{\pm}$, $\Omega^{\pm}$& 20 & $1.288  \pm  0.021$ & $0.582  \pm  0.009$ ($0.663  \pm  0.005$) & $0.035  \pm  0.008$ & $115.51  \pm  2.72$ ($93.12  \pm  0.19$) & 1.4 (1.8)\\
& 30 & $1.728  \pm  0.026$ & $0.603  \pm  0.006$ ($0.675  \pm  0.004$) & $0.101  \pm  0.013$ & $108.22  \pm  1.09$ ($95.84  \pm  0.17$) & 2.0 (2.2)\\
& 40 & $1.752  \pm  0.018$ & $0.615  \pm  0.004$ ($0.681  \pm  0.004$) & $0.079  \pm  0.011$ & $115.02  \pm  1.30$ ($98.87  \pm  0.13$) & 3.6 (3.6)\\
& 80 & $1.989  \pm  0.021$ & $0.602  \pm  0.005$ ($0.673  \pm  0.003$) & $0.058  \pm  0.008$ & $129.87  \pm  1.68$ ($106.54  \pm  0.12$) & 3.6 (3.4)\\
& 158 & $2.031  \pm  0.029$ & $0.610  \pm  0.003$ ($0.703  \pm  0.002$) & $0.083  \pm  0.007$ & $137.80  \pm  1.16$ ($109.24  \pm  0.11$) & 3.6 (3.4)\\

\hline
\end{tabular}
\caption{Summary of the fit results of $p_{\rm T}$-spectra of light and heavy strange hadrons after implementing the flow fluctuations with Gaussian distribution of transverse velocity, at different energies ranging from 20A to 158A GeV at SPS. The values $\eta_{max}$ are kept the same as no fluctuations scenario and adopted from Refs~\cite{Rode:2020vhu} and~\cite{Rode:2018hlj}. The corresponding fit results in no fluctuations scenario are quoted in parenthesis.}
\label{tabIII}
\end{table*}

In this section, we briefly introduce the non boost-invariant blast wave model. For more details, the reader is referred to Refs~~\cite{Dobler:1999ju,Rode:2018hlj,Rode:2020vhu}. Within the framework of this model, the single particle spectrum for central collisions with respect to transverse mass $m_T (\equiv\sqrt{p_T^2+m^2})$ and rapidity $y$ can be written as,
\begin{eqnarray}\label{therm}
  \frac{dN}{m_{T} dm_{T}  dy}
  &=&
  \frac{g}{2\pi}m_{T}\tau_F
  \int_{-\eta_{\max}}^{+\eta_{\max}} d\eta\,\cosh(y-\eta)
\nonumber\\
  &\times&
  \int_0^{R(\eta)} r_\perp\,dr_\perp \,
  \I_0\left(\frac{p_\T\sinh\rho(r_\perp)}{T}\right) \\
  &\times&
  \exp\left(\frac{\mu-m_\T\cosh(y{-}\eta)\cosh\rho(r_\perp)}
                 {T}\right) .
\nonumber
\end{eqnarray}
where $g$ is the degeneracy of particle species and $\eta$ ($\equiv\tanh^{-1}(z/t)$) is the space-time rapidity. Moreover, we have $\beta_{T}=\tanh(\rho)$ where $\rho$ is the flow rapidity in the transverse plane (or transverse rapidity) and $\beta_{T}$ is the collective transverse fluid velocity. Under the assumption that the common freeze-out of the fireball is instantaneous, the freeze-out time $\tau_F$ becomes independent of the transverse coordinate $r_{\perp}$ and occurs at kinetic freeze-out temperature $T$. Considering a Hubble like expansion of the fireball in the transverse plane, the transverse fluid velocity has radial dependence and is assumed to have the form:
\begin{equation}
\label{beta}
\beta_{T}(r_{\perp}) = \beta_{s}\left( {\frac{r_{\perp}}{R(\eta)}} \right).
\end{equation}
where $\beta_{s}$ denotes the transverse fluid velocity at the surface of the fireball. It is important to note that in the above equation, we have $R(\eta)$ in the denominator as opposed to $R_0$ in the model from Ref.~\cite{Dobler:1999ju}. Due to this characteristic, for a given non-zero $\eta$, the transverse flow goes to zero at the center and takes the maximum value $\beta_{s}$ at the edges of the fireball as $r_{\perp}$ approaches to $R(\eta)$. For the case of a linear parametrization, the average transverse flow velocity becomes $\langle\beta_T\rangle=\frac{2}{3}\beta_{s}$ and thus it is independent of $\eta$.

As discussed earlier, it is important to note that the presence of flow fluctuations has been neglected in the differential spectra shown in Eq.~\ref{therm}. Because of the finite system size, large fluctuations in the initial stage of the heavy-ion collisions may appear, even in the central collisions. These fluctuations can affect the initial conditions of the hydrodynamical expansion of the medium. Moreover, owing to the nonlinear nature of hydrodynamic equations, the event average of any hydrodynamical parameter is quite different from that for a smooth initial configuration. This leads to large differences in spectra obtained from hydrodynamical calculations with averaged initial conditions, compared to fluctuating initial conditions~\cite{Hama:2004rr, Andrade:2008xh, Hama:2009pk}. Therefore it is important to incorporate the collective flow fluctuations in blast-wave model to examine their effect on the kinetic-freeze-out parameters. To this end, we consider new form of non-boost-invariant blast-wave model averaged over an ensemble of the fluctuations, motivated from Ref~\cite{Akkelin:2009nx} for the transverse surface velocity, $\beta_{s}$ as:
\begin{eqnarray}\label{therm2}
  \frac{dN}{m_{T} dm_{T}  dy}
  &\propto&
  \frac{g}{2\pi}m_{T}\tau_F
  \int_{\beta^{\min}_{s}}^{\beta^{\max}_{s}} d\beta_{s}F(\beta_{s})
\nonumber\\
  &\times& 
  \int_{-\eta_{\max}}^{+\eta_{\max}} d\eta\,\cosh(y-\eta)
\nonumber\\
  &\times&
  \int_0^{R(\eta)} r_\perp\,dr_\perp \,
  \I_0\left[\frac{p_\T\sinh\rho(r_\perp)}{T}\right] \\
  &\times&
  \exp\left[\frac{\mu-m_\T\cosh(y{-}\eta)\cosh\rho(r_\perp)}
                 {T}\right] .
\nonumber
\end{eqnarray}

We consider two different profiles for distribution of $\beta_{s}$,

\begin{equation}\label{beta_fluc}
F(\beta_{s}) = \begin{cases}
1 &\text{:~Uniform}\\
\exp{\left[-\frac{(\beta_{s}-\beta^{0}_{s})^{2}}{\delta^{2}} \right]} &\text{:~Gaussian}
\end{cases}
\end{equation}

In first case, a flat or uniform distribution  of hydrodynamical velocities is considered with $\beta^{min}_{s}$ and $\beta^{max}_{s}$ being the lower and upper limit of the transverse flow velocities. The average of uniform distribution, defined as, $\beta^{0}_{s}$ $=$ ($\beta^{min}_{s}$ + $\beta^{max}_{s}$)/2. In the second case, a Gaussian distribution of hydrodynamical velocities is assumed with $\beta^{0}_{s}$ and $\delta$ being the mean and standard deviation, respectively. In this case, the lower and upper limit of the transverse flow velocities are taken to be 0 and 1, respectively. 

To make up for limited available incident energy, the freeze-out volume is restricted in the region $-\eta_{max}\le\eta\le\eta_{max}$, assuming the reflection symmetry about the center of mass. The transverse size is parameterized considering the elliptic shape of fireball in reaction plane, as follows,
\begin{equation}\label{ellipsoid}
R(\eta) = R_0\,\sqrt{1 - {\eta^2\over\eta_{\max}^2}}\,,
\end{equation}
where $R_0$ denotes the transverse size of the fireball at $\eta=0$. The dependence on $R_0$ factors out after changing the integral variable $r_\perp\to r_\perp/R$ in Eq.~\eqref{therm2} which lead to an overall factor of volume, $\tau_F R_0^2$. Moreover, the assumption of the boost-invariance is relaxed by the explicit dependence of system boundary in the transverse plane on the longitudinal coordinate, as parameterized in Eq.~\eqref{ellipsoid}. At the freeze-out surface, the temperature is assumed to be constant. Moreover, the transverse flow gradient is independent of $r_{\perp}$ and it has only $\eta$ dependence through $R(\eta)$. One can notice from Eq.~\eqref{therm2} that the variable $r_{\perp}$ takes values between $0 \le r_{\perp} \le R(\eta)$. However, the transverse velocity $\beta_{T}(r_{\perp})$ given in Eq.~\eqref{beta} remains finite and lies in the physical range (preserves causality) for $\beta_{s}<1$, even though $R(\eta) \rightarrow 0$ as $\eta \rightarrow \pm{\eta_{max}}$. In addition, we observe from Eq.~\eqref{beta} that the transverse flow gradient along $r_{\perp}$ diverges as $\eta \rightarrow \pm{\eta_{max}}$. This makes the model unsuitable for those analyses involving the quantities which depend on gradients, such as dissipative effects. Nevertheless, in our framework we do not deal with such gradients since we are employing the non-dissipative blast wave model and hence such issues are not encountered in the implementation of this model.

One may argue that because of the elliptic shape of the
fireball in the reaction plane in Eq.~{\eqref{ellipsoid}}, the system size decreases away from midrapidity and hence flow fluctutations may get stronger. Therefore, the parameters in Eq.~{\eqref{beta_fluc}} should depend on space-time rapidity. However, it is important note that almost all of the analyzed species in the present work are measured at mid-rapidity. Moreover, the $\eta$-dependence of temperature was studied in Ref.~\cite{Dobler:1999ju}. The authors concluded that the change in the temperature due this dependence is small without compromising the $\chi^{2}/\rm Ndf$, which suggests that same might also be true for transverse velocity. Therefore the effect of $\eta$-dependence on the model parameters is expected to be small and hence is ignored in the present work.
\begin{figure*}[t]
\includegraphics[scale=0.35]{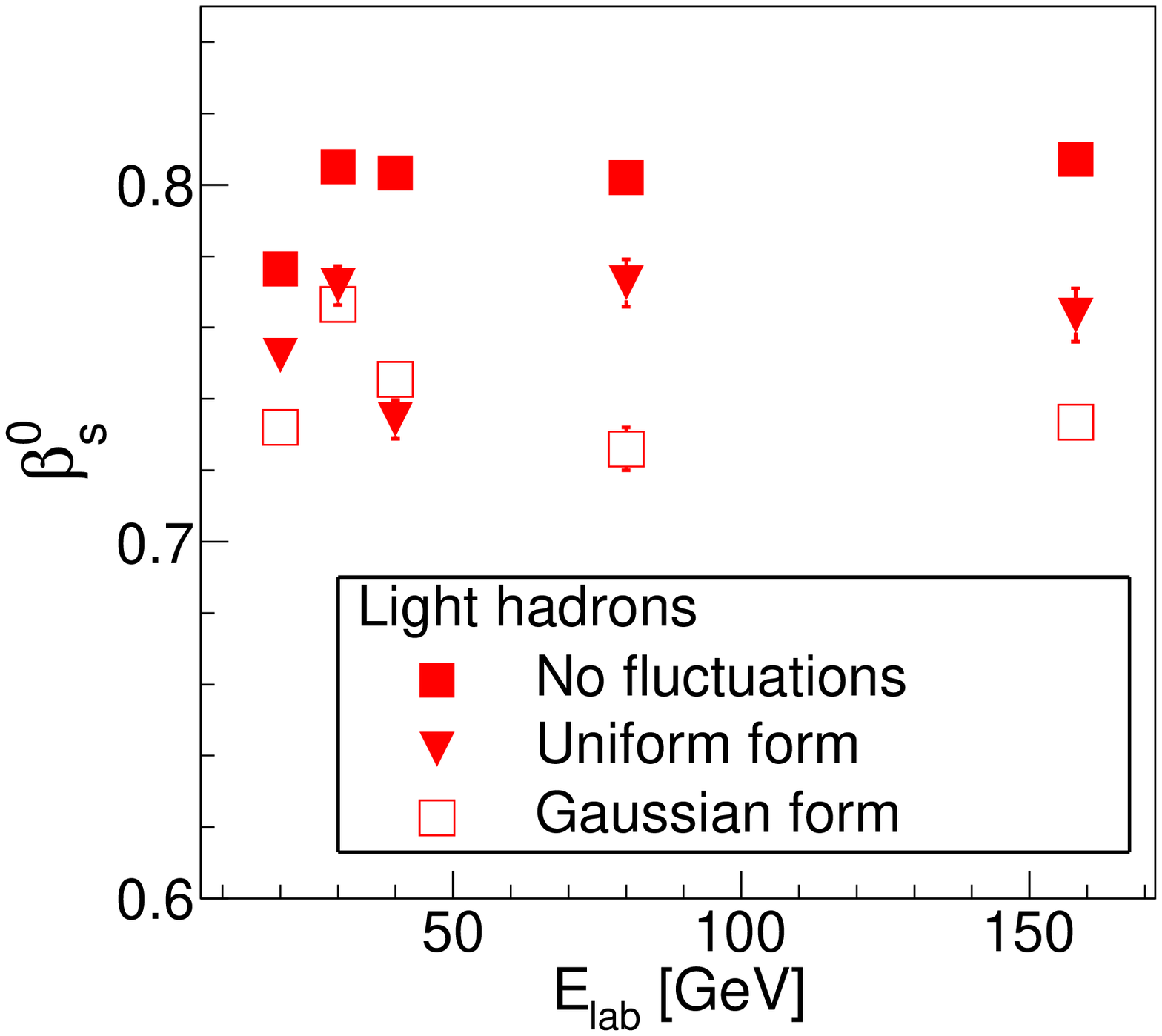}
\includegraphics[scale=0.35]{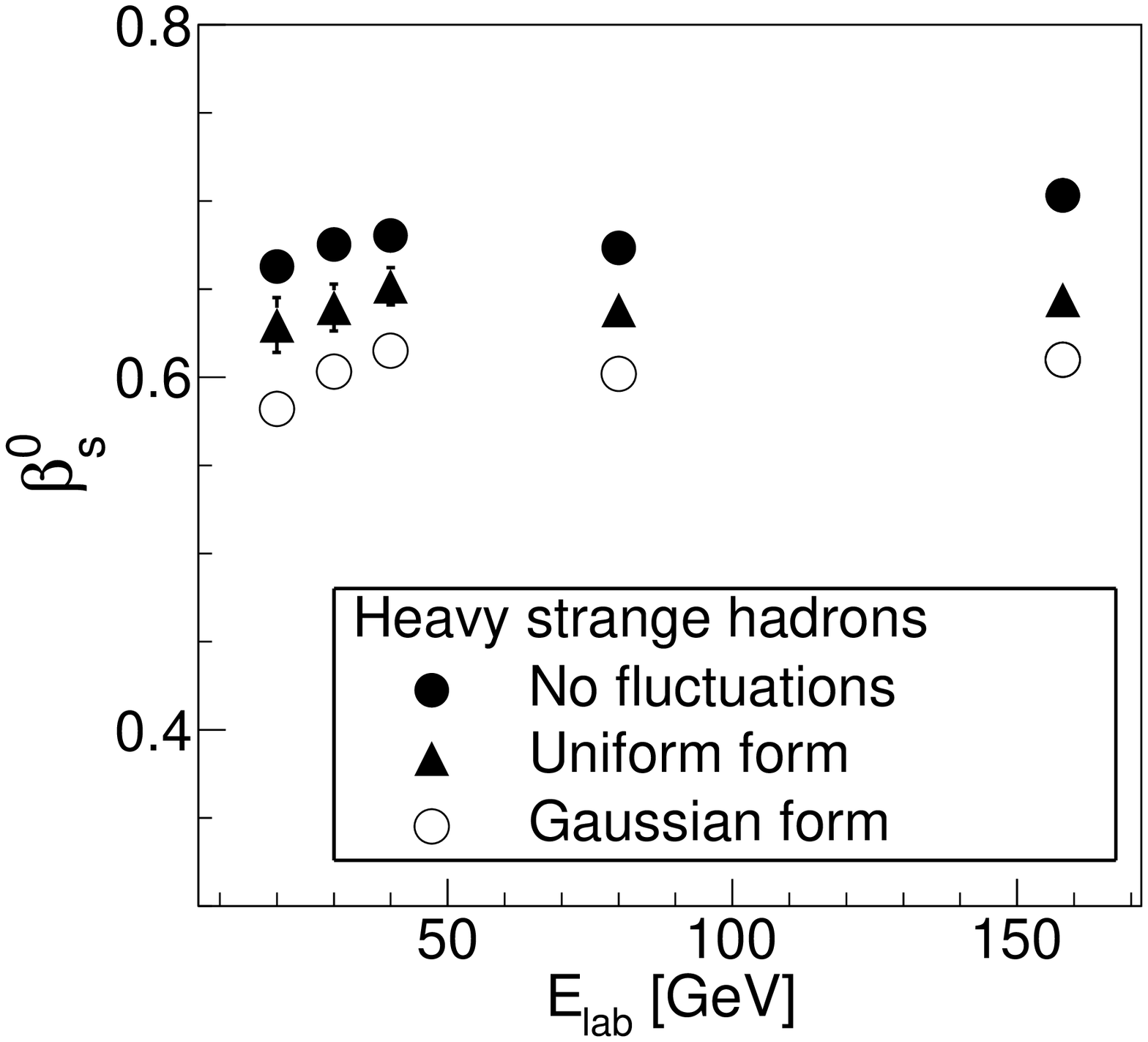}\\
\includegraphics[scale=0.35]{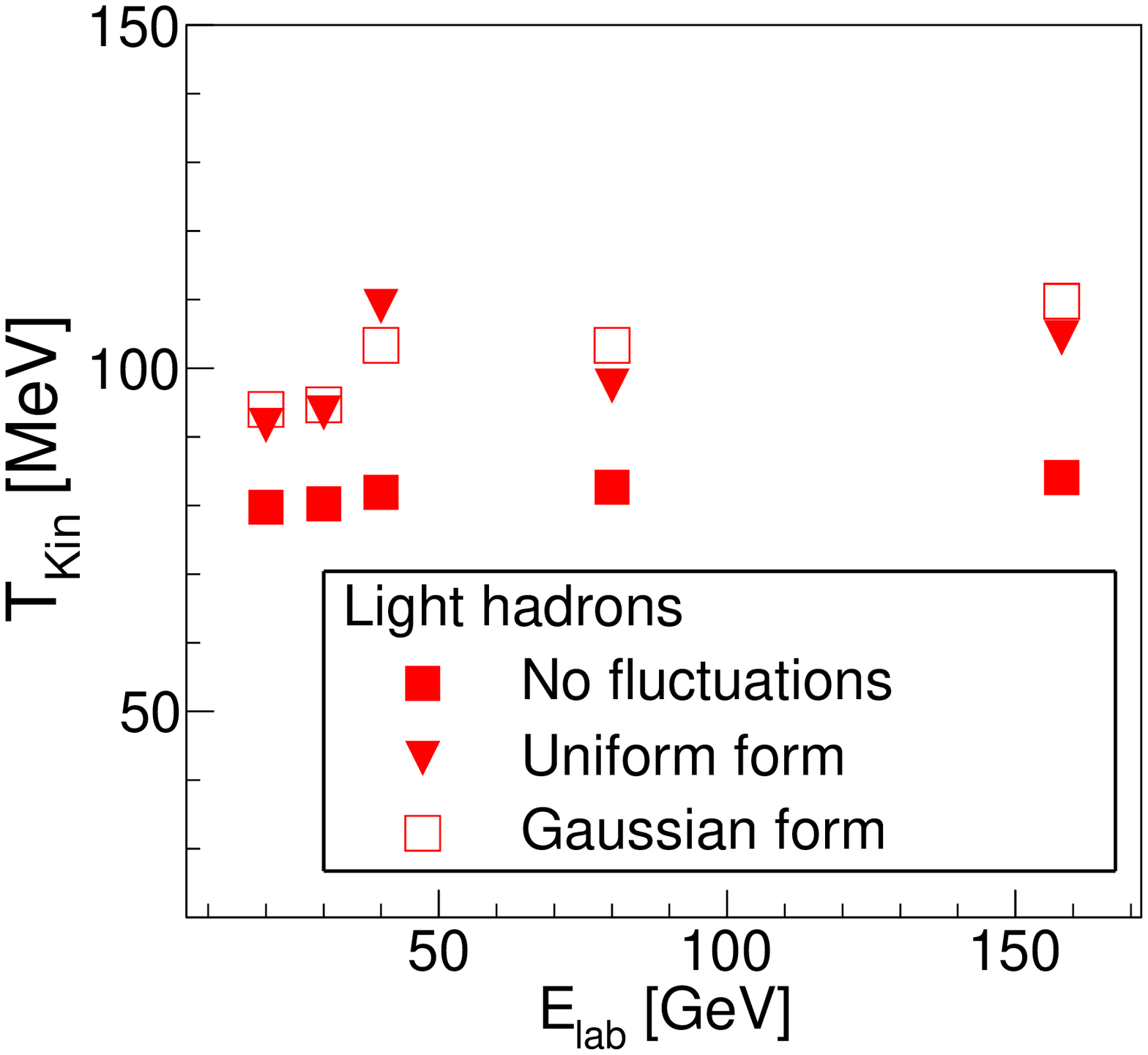}
\includegraphics[scale=0.35]{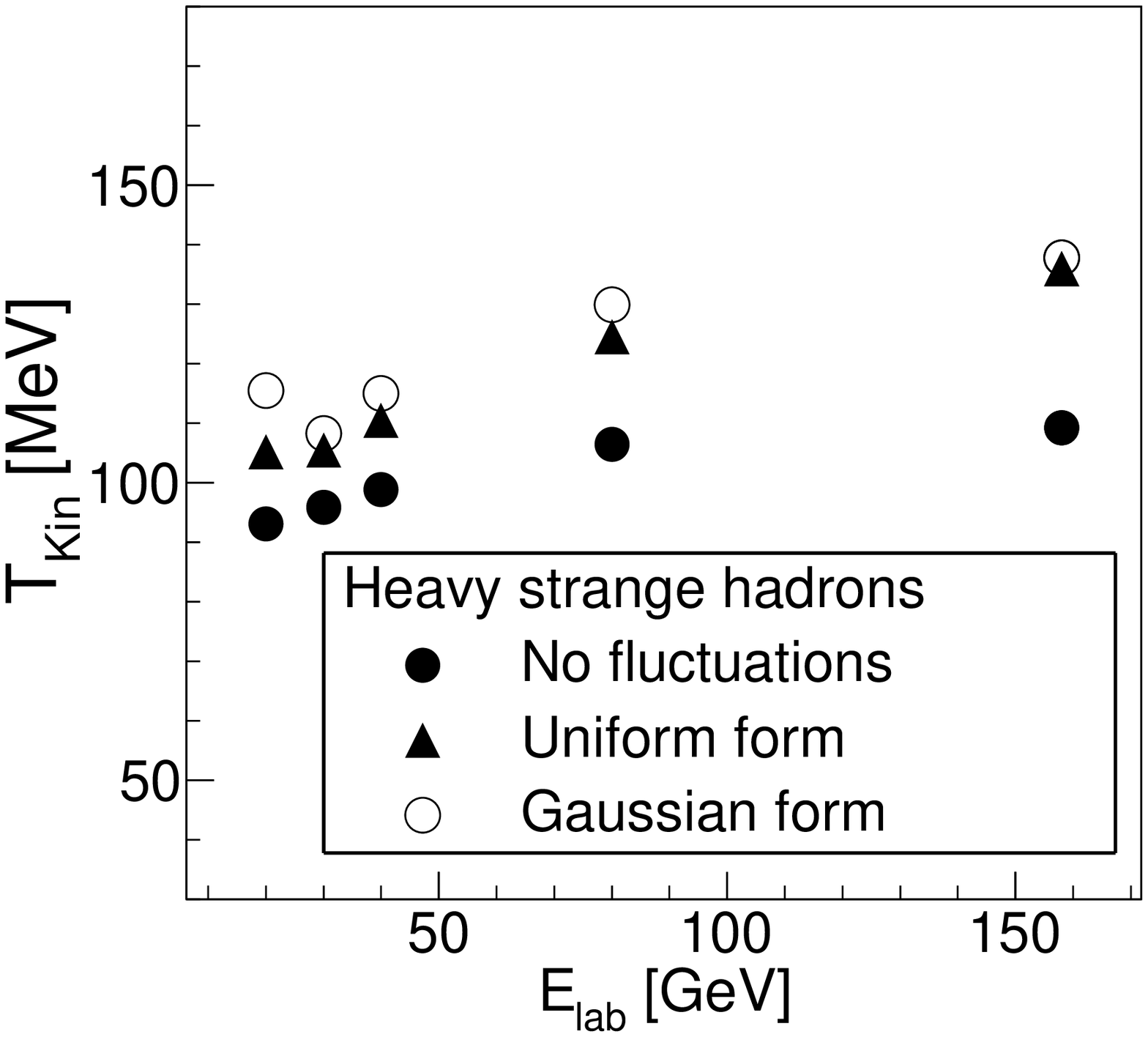}
\caption{Variation of the $\beta^{0}_{s}$ (top two plots) and $T_{kin}$ (bottom two plots) for heavy strange and light hadrons with incident beam energy ($\rm E_{\rm lab}$). $\beta^{0}_{s}$ estimated for the case of uniform fluctuations is obtained by taking the mean of $\beta^{min}_{s}$ and $\beta^{max}_{s}$. Visible vertical bars are associated errors on the parameters and for the rest of the parameters, errors are within the marker size.}
\label{fig6}
\end{figure*}

%

%


\section{Results and discussions}

Results obtained from this study have been presented and discussed in this section. For this purpose, we have analyzed the measured transverse momentum spectra ($p_{T}$) of light, heavy strange and charmed (only at $\rm E_{Lab}=158A $ GeV) hadrons produced in central Pb--Pb collisions from NA49 and NA50 collaboration~\cite{Alt:2008qm, Alt:2008iv, Alt:2004kq} at SPS in the beam energy range $\rm E_{Lab}=20A-158A $ GeV. The hadrons analyzed in this manuscript were categorized according to their masses, following the intuition that heavy particles may decouple earlier than lighter ones. Note that we have focused only at SPS energies as the data for heavier particles in the desired kinematic regions are barely available at lower beam energies. Resonance decay contributions to the lightest hadron in our dataset, i.e. pions, are taken into consideration following the formalism in Ref.~\cite{reso-decay}. All $p_{T}$-spectra analyzed here are calculated at center of measured rapidity region (e.g. at $y_{c.m.}=0.1$, for $0<y_{c.m.}<0.2$) of the hadron. We have explicitly checked that fit of $p_{\rm T}$-spectra at the central value of measured rapidity coverage of a particular species does not yield any significant changes in the values of the parameters when compared with the parameters obtained by fitting the spectra integrated over the measured rapidity region. Therefore, the main message of our paper remains unaltered. The fits of $p_{T}$-spectra are performed simultaneously for each category of hadrons by minimizing the value of $\chi^{2}/N_{dof}$, where $N_{dof}$ is the number of degrees of freedom defined as the number of data points minus the number of fitting parameters. In our analysis, the minimization procedure was performed using MINUIT~\cite{minuit2} package available in ROOT framework~\cite{root}.

\begin{figure*}[ht]
\includegraphics[scale=0.35]{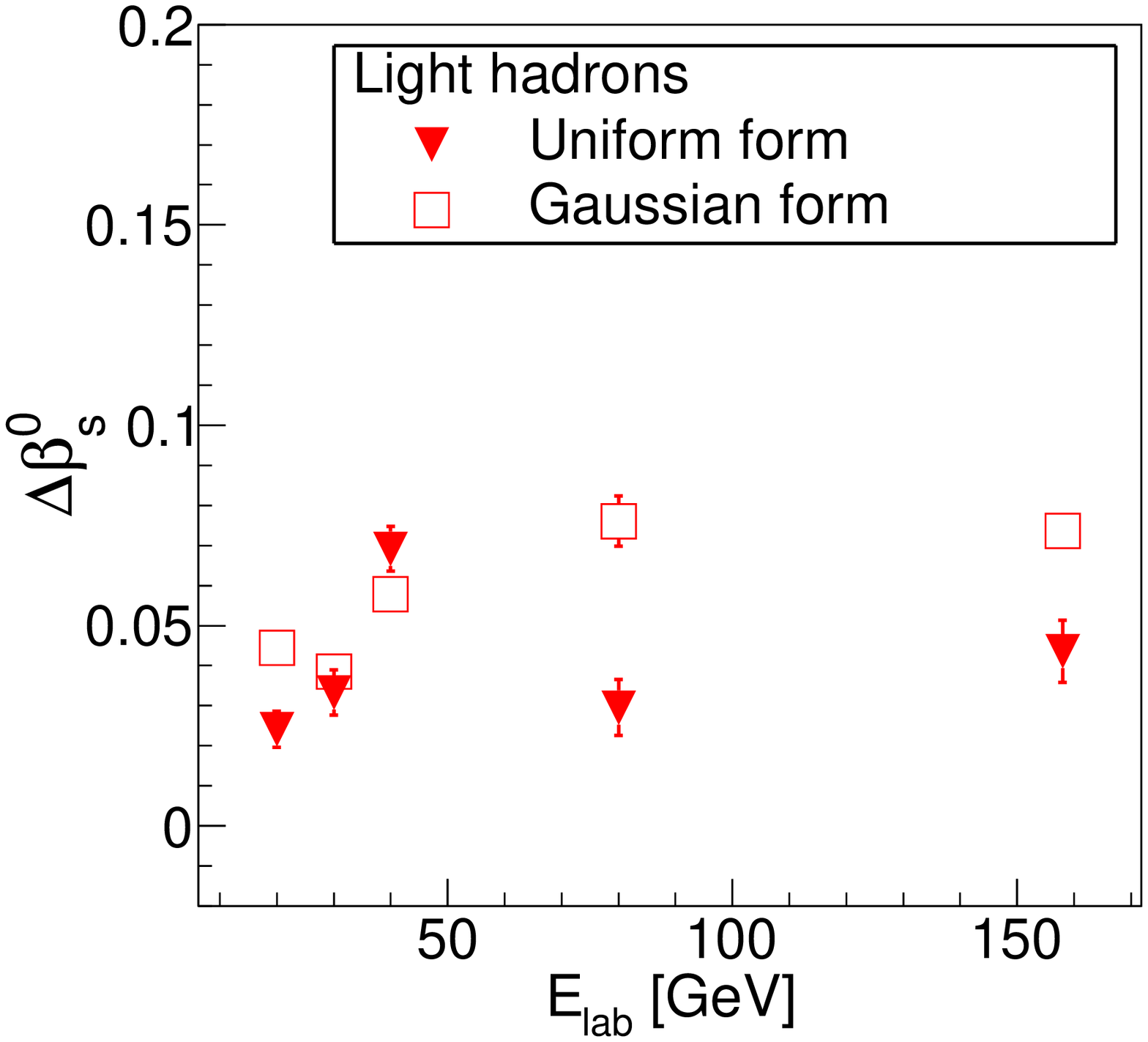}
\includegraphics[scale=0.35]{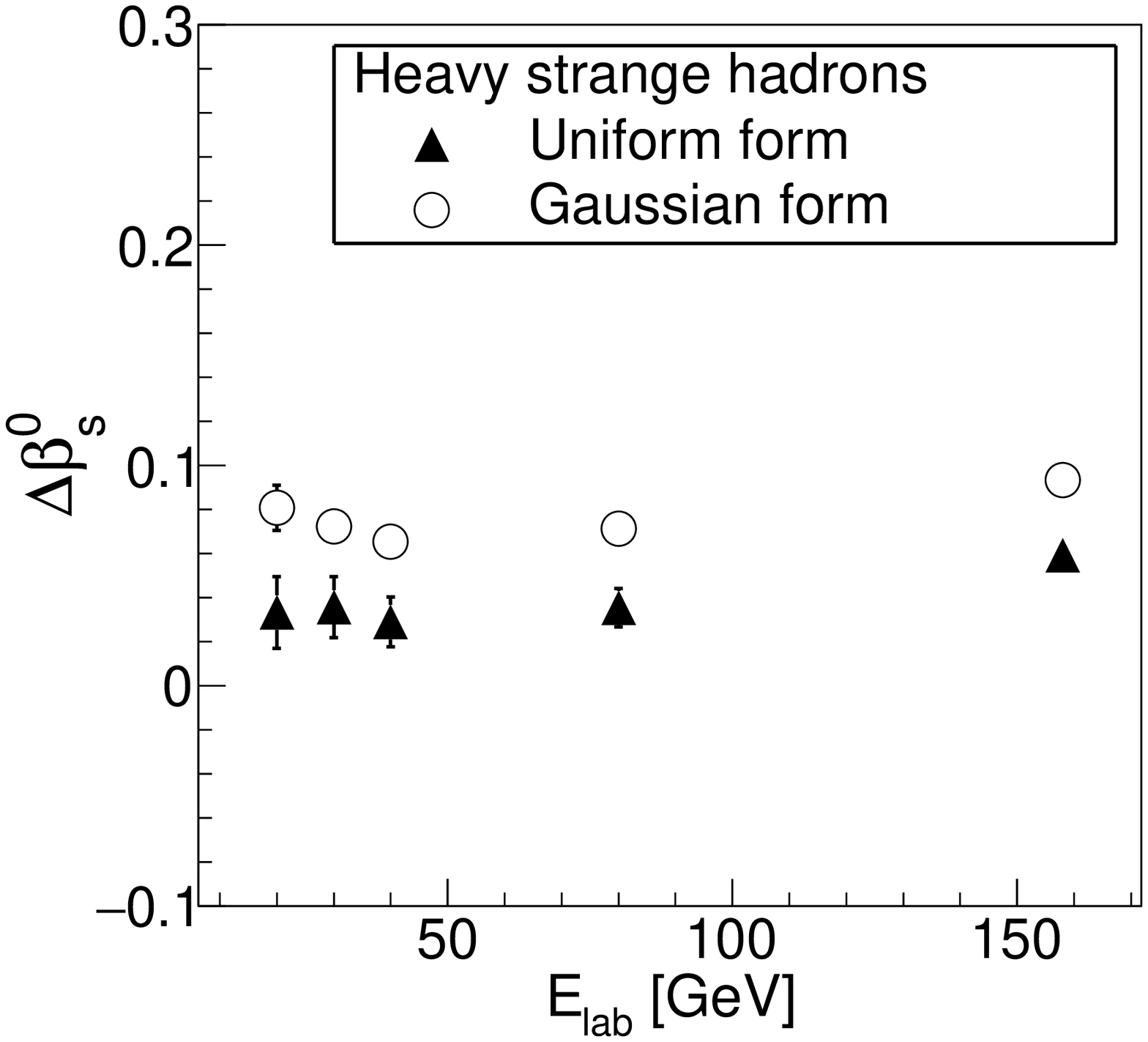}\\
\includegraphics[scale=0.35]{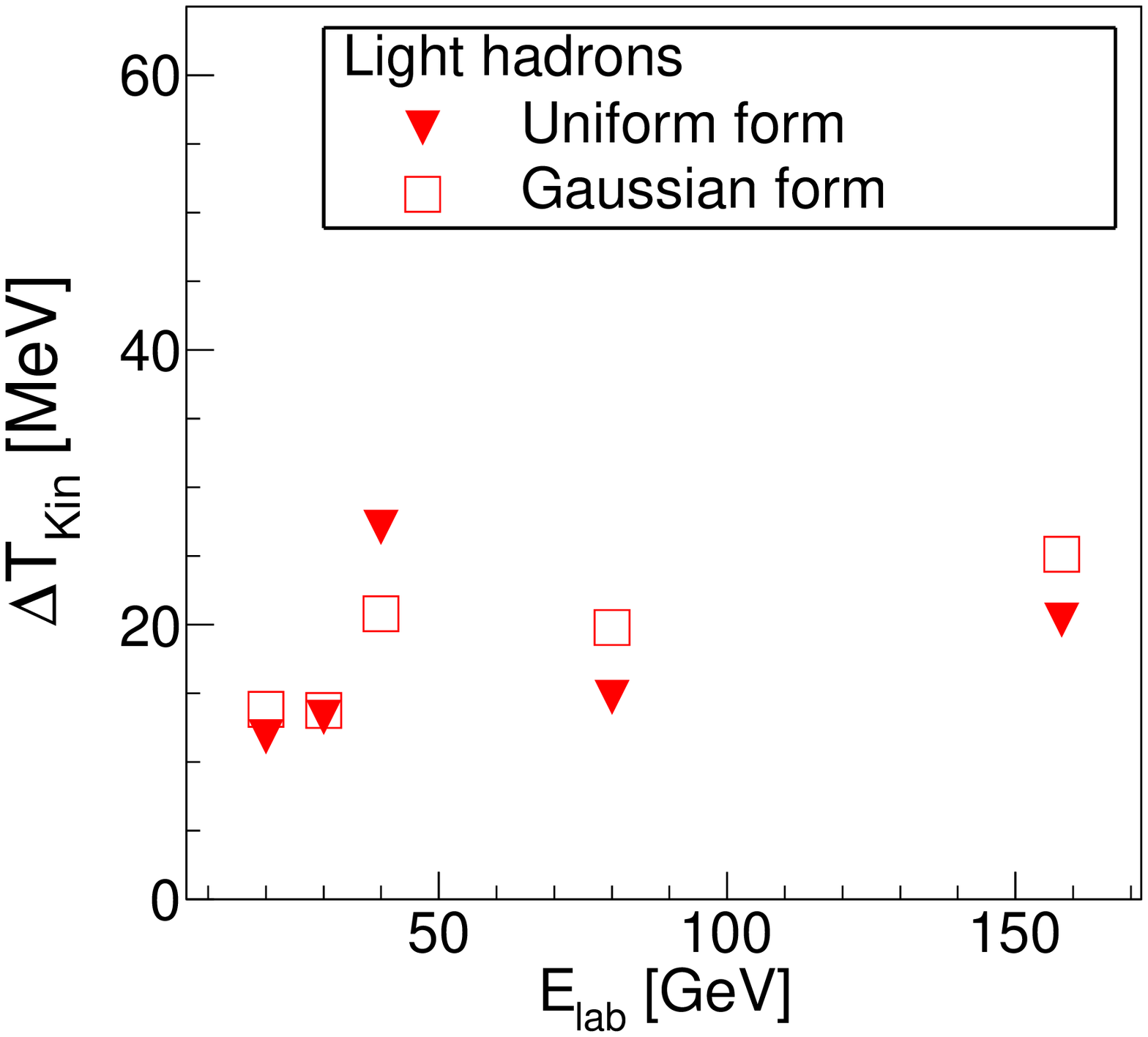}
\includegraphics[scale=0.35]{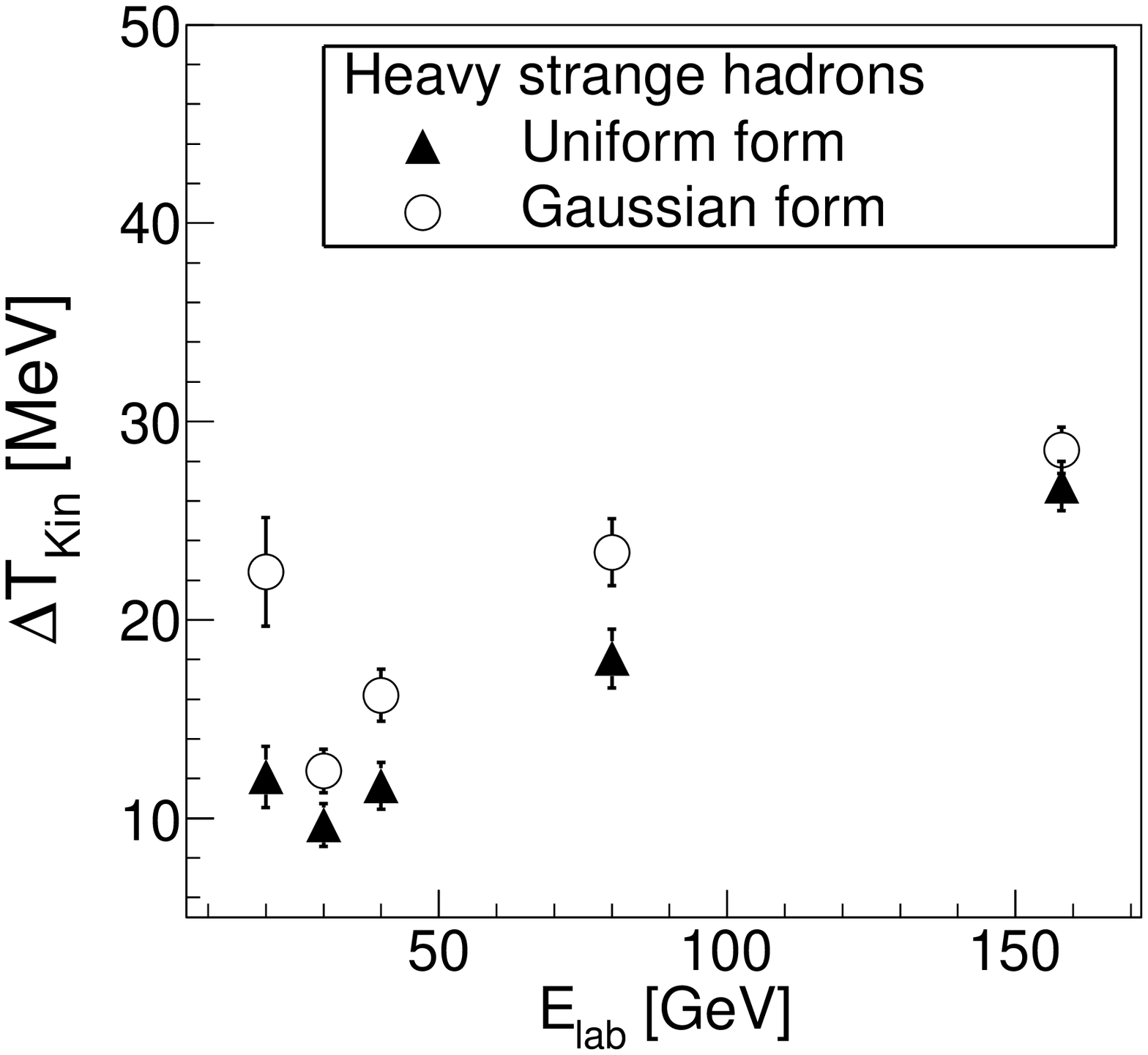}
\caption{Difference of $T_{kin}$ ($\Delta T_{kin}=|T_{kin}^{\rm UF/GF}-T_{kin}^{\rm NF}|$) and $\beta^{0}_{s}$ ($\Delta \beta^{0}_{s}=|{\beta^{0}_{s}}^{\rm NF}-{\beta^{0}}_{s}^{\rm UF/GF}|$) of light and heavy strange hadrons for both uniform and Gaussian form with respect to no fluctuations scenario as a function of beam energy. Vertical bars are propagated errors after the subtraction.}
\label{Delta_T_HS}
\end{figure*}

For the adopted linear transverse flow profile, essentially, there are three parameters associated with our non-boost invariant blast-wave model; see Eq.~(\ref{therm}). These parameters are $T_{kin}$, $\eta_{max}$ and $\beta_{s}$ out of which two parameters, $T_{kin}$ and $\beta_{s}$ are sensitive to the $p_T$-spectra. On the other hand, the $p_T$-spectra is rather insensitive to $\eta_{max}$ and the rapidity spectra is insensitive to the other two parameters, $T_{kin}$ and $\beta_{s}$. Moreover, because of the non-Bjorken flow, the rapidity spectra are sensitive to $\eta_{max}$~\cite{Rode:2018hlj, Du:2023gnv}. This was the main reason that we did not consider to analyze rapidity spectra of the hadrons under study using Eq.~(\ref{figs1}). Along with this, we also checked this explicitly by fitting the rapidity spectra by intergating Eq.~(\ref{figs1}) with respect to $p_{T}$ to obtain the desired longitudinal spectra and we found that the parameter, $\eta_{max}$ is unchanged. Therefore, we have used the values of $\eta_{max}$ obtained from our previous analyses~\cite{Rode:2018hlj,Rode:2020vhu}, where, the values of $\eta_{max}$, $T_{kin}$ and $\beta_{s}$ were obtained recursively. Firstly, $\eta_{max}$ is fixed from the simulatenous fits of the rapidity distributions with initial guess of $T_{kin}$ and $\beta_{s}$, and then this $\eta_{max}$ is used to fit corresponding $p_{T}$ distributions. These newly extracted $T_{kin}$ and $\beta_{s}$ values are then used to get updated $\eta_{max}$. This procedure converges rather quickly.

As discussed in Section II, there are two cases, namely, uniform and Gaussian distributions, corresponding to the form of fluctuations $F(\beta_{s})$, considered in this study. First, we start by fitting the $p_{T}$-spectra of light and heavy strange hadrons using Eq.~(\ref{therm2}) with the former case, $F(\beta_{s}) = 1$. Using this approach we extract three parameters, namely, $\beta^{min}_{s}$, $\beta^{max}_{s}$ and $T_{kin}$. For the comparison, the average of the uniform dostribution is also estimated which is defined as, $\beta^{0}_{s}$ $=$ ($\beta^{min}_{s}$ + $\beta^{max}_{s}$)/2. The quality of the fits is better than and in few cases similar to the default case, i.e. no fluctuations scenario. 

The obtained fit parameters are tabulated in Table~\ref{tabII}.  We have noticed that the predicted new $T_{kin}$ values are higher than the ones from our previous analyses, where it was between $80-85$ MeV for light hadrons and $90-110$ MeV for heavy strange hadrons.  This seems to be the consequence of the implementation of the flow fluctuations into our model, to which the initial hydrodynamical conditions are expected to be sensitive and subsequently, affect the kinetic-freezeout conditions. 
It is important note that for uniform flow fluctuation profile, the fitted value of transverse flow velocity is smaller than the default case, i.e., without fluctuations. In order to compensate for smaller transverse flow velocity, the fitted value of kinetic freeze-out temperature is larger than the default case. This may be attributed to the fact that both transverse flow velocity and temperature leads to hardening of transverse momentum spectra

Moving on to the second case of flow fluctuations, we have used the Gaussian description of hydrodynamical velocities (Eq.~(\ref{beta_fluc})) and have fitted the $p_{T}$-spectra of light and heavy strange hadrons using Eq.~(\ref{therm2}). In this case, we have fixed the lower and upper limits of the Gaussian function, $F(\beta_{s})$ to be 0 and 1, respectively. However, the parameters $T_{kin}$, $\delta$ and $\beta^{0}_{s}$ are kept as free. Here as well, the quality of the fits is better, and in a few cases similar to the no fluctuations scenario. The fit parameters obtained from this analysis are tabulated in Table~\ref{tabIII}. The observation is that the $T_{kin}$ values are even higher than uniform description case and $\beta^{0}_{s}$ values are smaller than the ones from our previous analyses, where it was between $0.77-0.82$ for light hadrons and $0.65-0.70$ for heavy strange hadrons. Moreover, values of $\delta$ parameter vary between $0.03-0.15$ for both light hadrons and heavy strange hadrons. 

As mentioned earlier, in case of Gaussian flow fluctuation profile the transverse flow velocity is even smaller than uniform flow fluctuation profile. Therefore, in order to compensate for smaller transverse flow velocity, the fitted value of kinetic freeze-out temperature is larger than the uniform case. One may argue that because of the fixed limits on the Gaussian distribution, i.e., 0 and 1, the tails may cut asymmetrically and subsequently, the central point of the Gaussian distribution may not be correct parameter, instead, the average $\beta_{s}$ may be better parameter to compare with Uniform profile scenario. However, we have checked that the average and the central values agree upto three decimal places, and therefore, it makes no significant difference to the results.

\begin{figure*}[t]
\begin{picture}(220,200)
\put(0,0){\includegraphics[scale=0.4]{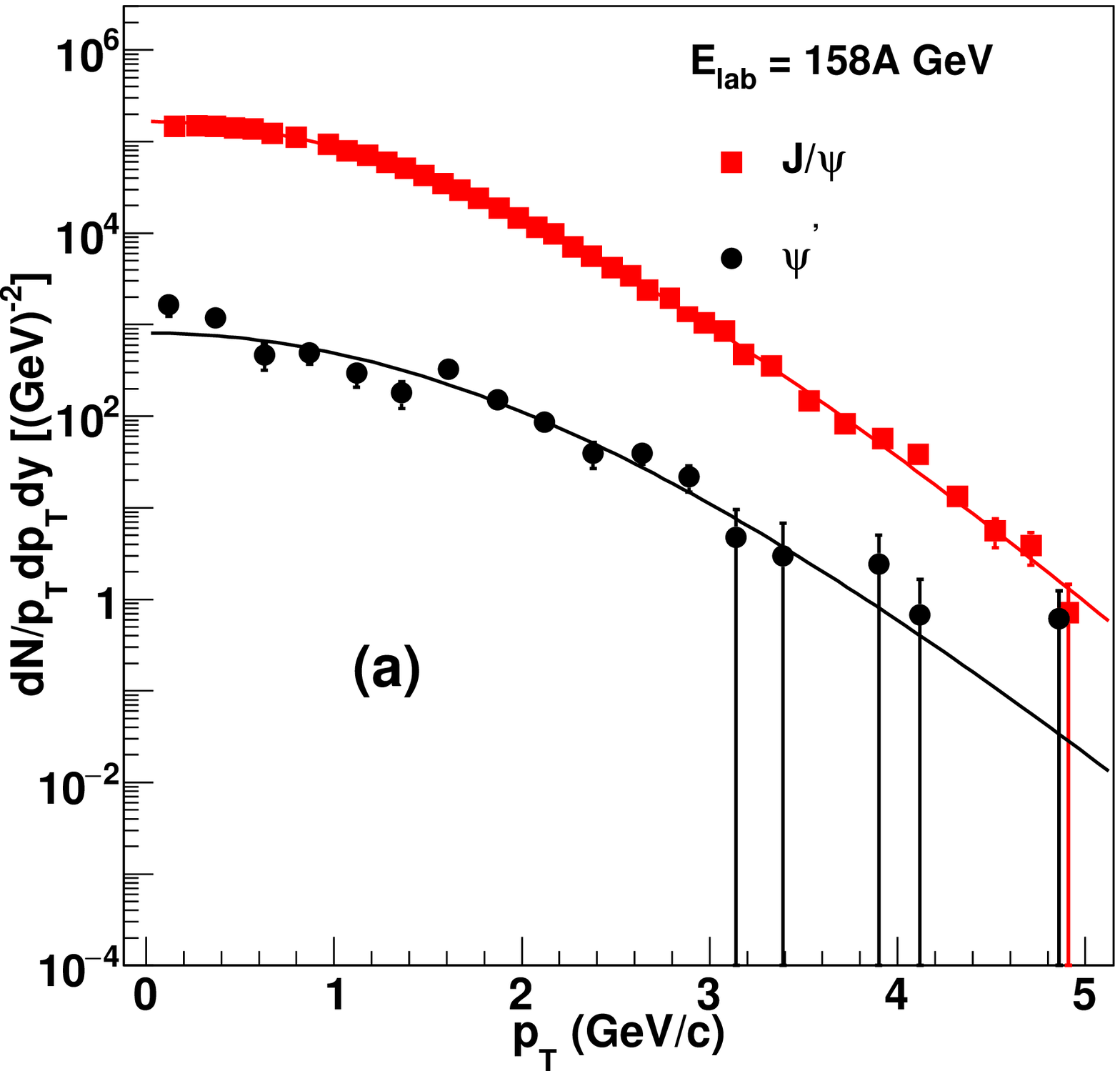}}
\end{picture}
\begin{picture}(200,200)
\put(0,0){\includegraphics[scale=0.4]{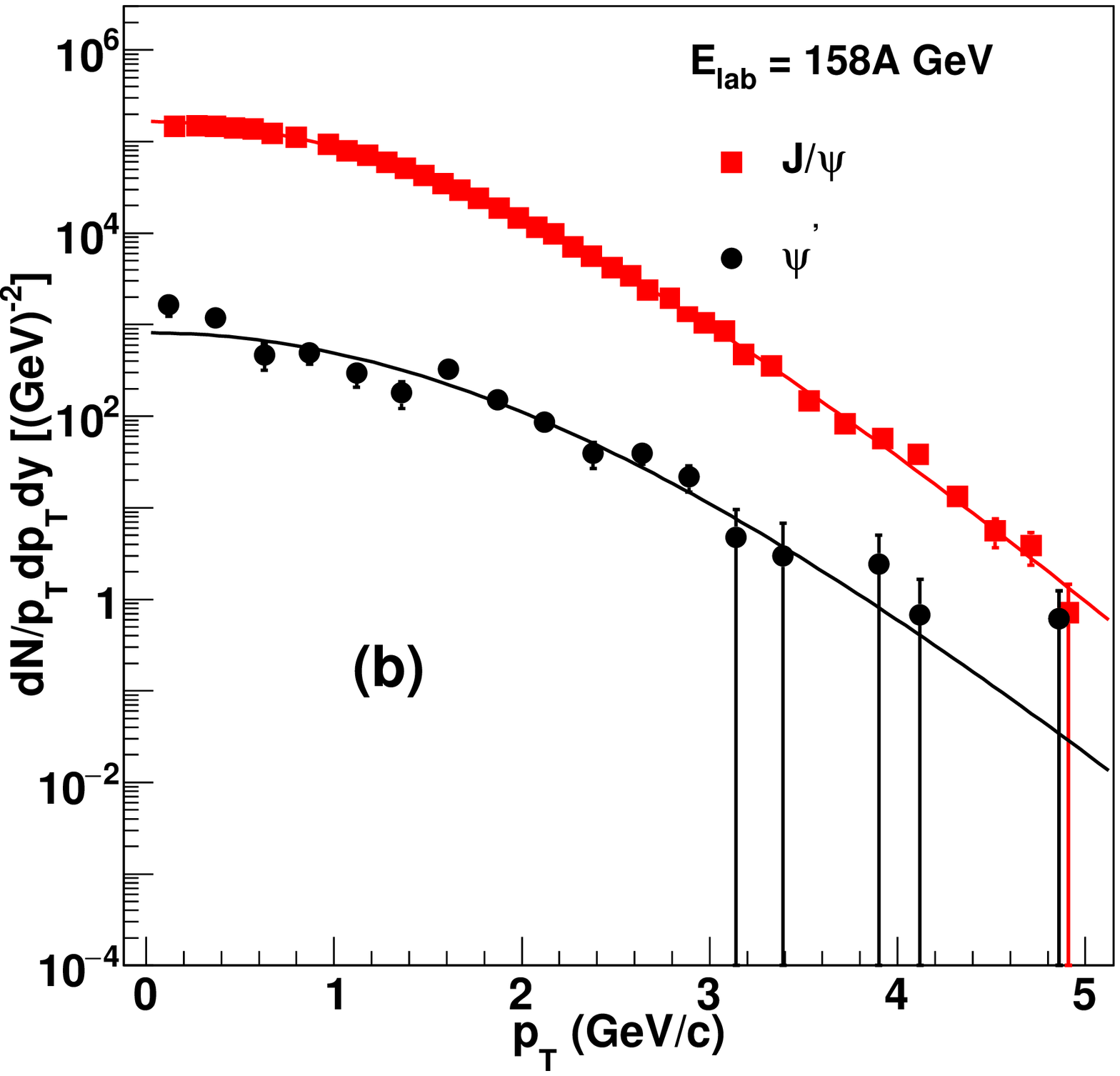}}
\end{picture}

\caption{Simultaneously fitted $p_{T}$-spectra of $\rm J/\psi$ and $\psi^{'}$ at 158A GeV beam energies using (a) Uniform and (b) Gaussian description of transverse flow fluctuations. Error bars indicate available statistical error.}
\label{fig3}
\end{figure*}

Next we look at the beam energy dependence of these extracted fit parameters as shown in Fig.~\ref{fig6}. Here, we have compared the values of $\beta^{0}_{s} $ and $T_{kin}$ of light hadrons and heavy strange hadrons obtained from this study with no fluctuations scenario. Now, variation in the values of these parameters can be clearly seen in case of flow fluctuation with respect to no fluctuations. It is also interesting to observe even stronger beam energy dependence of $T_{kin}$ in case of the flow fluctuations. Moreover, looking at the excitation functions of these parameters qualitatively, one may find the trends interesting. To investigate this in detail, we estimate the differences of $T_{kin}$ and $\beta^{0}_{s}$ with respect to no fluctuations case, $\Delta T_{kin}=|T_{kin}^{\rm UF/GF}-T_{kin}^{\rm NF}|$ and $\Delta \beta^{0}_{s}=|{\beta^{0}_{s}}^{\rm NF}-{\beta^{0}}_{s}^{\rm UF/GF}|$ as a function of beam energies as shown in Fig.~\ref{Delta_T_HS}. These quantities have a non-monotonous structure for both uniform as well as Gaussian formulations, with a minima/maxima around $\rm E_{Lab}\approx30A-40A$ GeV which is an interesting beam energy region. There has been many instances in the beam energy domain, $\rm E_{Lab}=20A-158A$ GeV where various observables have shown some interesting irregularities around  $\rm E_{Lab}\approx30A-40A$ GeV~\cite{Bleicher:2011jk}. This behaviour has often been linked to the potential signature of the onset of deconfinement. However, in our case, one needs to be careful and perform more detailed investigations to make any robust claims.

Moving on, charmonia, i.e. J/$\psi$, $\psi'$ have been analyzed in boost-invariant scenario~\cite{Gorenstein} and in non-boost-invariant case as well~\cite{Rode:2020vhu}, based on the hypothesis that the production of these hadrons happened through statistical coalescence and further freeze-out during hadronization. In the present study, after light and heavy strange hadrons, similar exercise was performed for $J/\psi$ and $\psi^{'}$~\cite{Abreu:2000xe} at $\rm E_{Lab}=158A $ GeV. Note that same $\eta_{max}$ value ($=1.70$) from our previous study was used for the fits. Similar fit qualtity as our previous analysis has been achieved here as well for both uniform and Gaussian distribution of the transverse velocities.  The values of the parameters in uniform distribution case are, $\beta^{min}_{s}=0.24$, $\beta^{max}_{s}=0.36$ and $T_{kin}=164$ MeV. For Gaussian flow fluctuations we obtain, $T_{kin}=165$ MeV, $\delta=0.05$ and $\beta^{0}_{s}=0.3$. Interestingly enough, the values of both $T_{kin}$ and $\beta^{0}_{s} $ are found to be similar to the case of no fluctuations, which was, $T_{kin}=164$ MeV and $\beta^{0}_{s}=0.3$. Moreover, the $\beta_{s}^{0}$ and its spread are lower than those of light and heavy strange hadrons where the freeze-out parameters, $T_{kin}$ and $\beta^{0}_{s}$, showed sensitivity to the assumption of flow fluctuations.

\begin{figure}[ht]
\includegraphics[scale=0.44]{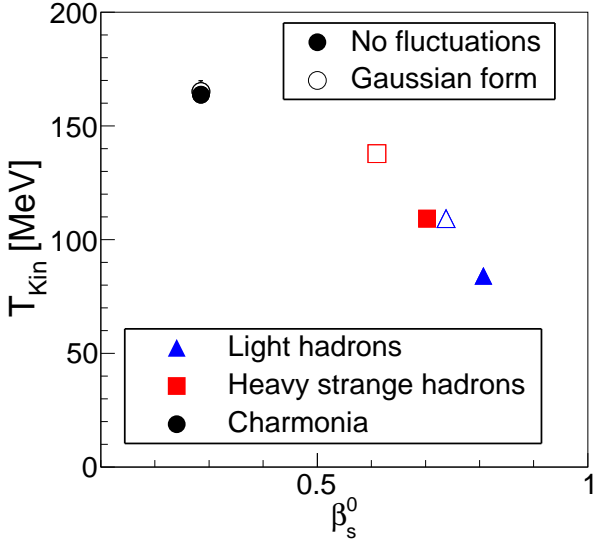}
\caption{The (partial) expansion history of the fireball created in $158A$ GeV central Pb--Pb collisions. The points indicate the temperature ($T_{kin}$) and transverse collective flow velocity ($\beta^{0}_{s}$) of the system at the time of light hadron kinetic freeze-out (filled triangle), heavy strange kinetic freeze-out (filled square) and charm kinetic freeze-out (filled circle). The values corresponding to Gaussian form of fluctuations are shown in empty symbols. Errors on the parameters are within the marker size.}
\label{fig7}
\end{figure}

In Fig~\ref{fig7}, an updated partial expansion history of the fireball after incorporation of the flow fluctuations into the transverse momentum spectra is presented. The freeze-out parameters for light, heavy strange and charmed hadrons obtained with no fluctuations and Gaussian prescription are plotted at $\rm E_{Lab}=158A $ GeV. It is very interesting to see that the effect of flow fluctuations on the freeze-out parameters for charmed hadrons is quite small compared to other two groups of species. This can be interpreted as follows: Due to small rescattering cross-sections in the hadronic phase, the momentum distributions of charmonia are also frozen near the phase boundary, similar to their chemical composition closer to the hadronization. This reflects in the fact that $T_{kin}$ for charmonia is close to $T_{c}$ or $T_{CFO}$. Because of this, the radial flow and associated fluctuations are not fully developed and show insensitivity as opposed to heavy strange and light hadrons.  Our study may provoke more efforts in this direction as well as subsequent studies from us will be performed in due time.


\section{Summary}

To summerize, we have made some efforts to study the effect of flow fluctuations on the kinetic freeze-out parameters of various particle species. For this purpose, we have modified the non-boost-invariant blast-wave model following Ref.~\cite{Akkelin:2009nx} where the authors incorporated the flow fluctuations into the boost-invarint blast-wave model. Two different functional forms of the $\beta_{s}$ distribution were considered, namely uniform and Gaussian description. We analyzed the transverse momentum spectra of different hadron species in central Pb--Pb collisions at different SPS beam energies. The transverse momentum spectra were fitted simultaneously to obtain various freeze-out parameters such as $\beta^{0}_{s} $ and $T_{kin}$. The temperatures obtained using both descriptions showed higher values compared to the case of no fluctuations scenario where it was between $80-85$ MeV for light hadrons and $90-110$ MeV for heavy strange hadrons. With inclusion of flow fluctuations , the temperature varies between $90-110$ MeV for light hadrons and $105-140$ MeV for heavy strange hadrons. Similarly, decrease in the $\beta^{0}_{s} $ values was observed for both descriptions at all beam energies with respect to no fluctuations scenario where it was between $0.77-0.82$ for light hadrons and $0.65-0.70$ for heavy strange hadrons. Incorporation of fluctuations made $\beta_{s}$ to reduce between  $0.73-0.76$ for light hadrons and $0.58-0.64$ for heavy strange hadrons. Moreover, we saw a stronger increase in the temperature as function of beam energies compared to no fluctuations scenario. Furthermore, values of standard deviation, i.e. $\delta$ parameter varies between $0.03-0.15$ for both light hadrons and heavy strange hadrons. We also fitted the charmonia at $E_{lab}=158A$ GeV and found that the temperature as well as $\beta^{0}_{s} $ values remain almost unchanged for both descriptions with respect to no fluctuations scenario. This suggests that the incorporation of flow fluctuations does not affect the kinetic freeze-out conditions for charmonia. This could be due to the fact that the radial flow and corresponding fluctuations are not fully developed due to freezing of momentum spectra immediately after or simultaneously at chemical freeze-out and therefore, the parameters are robust against the flow fluctuations. This is one of the interesting findings of our work.

As an outlook, these results can trigger further attempts to look at the flow fluctuations more closely at different centralities and also in the explanation of anisotropic flow coefficients. When the experimental measurements of anisotropic flow coefficients of identified hadrons including charmonia with various cumulants becomes available, the findings of this model can be verified. Moreover, it will be interesting to repeat such exercise with charmed hadrons for lower energy collisions, when the data become available. This can be achieved with the upcoming measurements at SPS and we leave this analysis for the future.

%
\begin{acknowledgements}
A.J. is supported in part by the DST-INSPIRE faculty award under Grant No. DST/INSPIRE/04/2017/000038.
\end{acknowledgements}
%

\end{document}